\documentclass[sigconf]{acmart}
\AtBeginDocument{%
  \providecommand\BibTeX{{%
    \normalfont B\kern-0.5em{\scshape i\kern-0.25em b}\kern-0.8em\TeX}}}

\usepackage{xspace}
\usepackage{hyperref}
\usepackage{multirow}
\usepackage{xcolor}
\usepackage{colortbl}
\usepackage{graphicx}
\usepackage{booktabs}
\usepackage{makecell}
\usepackage{soul}
\usepackage{setspace}

\def\clochat{CloChat\xspace}

\definecolor{mypurple}{RGB}{179, 102, 255}

\copyrightyear{2024}
\acmYear{2024}
\setcopyright{rightsretained}
\acmConference[CHI '24]{Proceedings of the CHI Conference on Human Factors in Computing Systems}{May 11--16, 2024}{Honolulu, HI, USA}
\acmBooktitle{Proceedings of the CHI Conference on Human Factors in Computing Systems (CHI '24), May 11--16, 2024, Honolulu, HI, USA}
\acmDOI{10.1145/3613904.3642472}
\acmISBN{979-8-4007-0330-0/24/05}

\begin{document}

\title{\clochat: Understanding How People Customize, Interact, and Experience Personas in Large Language Models}


\author{Juhye Ha}
\affiliation{
  \institution{Graduate School of Information \\Yonsei University}
  \city{Seoul}
  \country{Korea}
}
\email{juhye0329@yonsei.ac.kr}

\author{Hyeon Jeon}
\affiliation{
  \institution{Dept. of CSE \\Seoul National University}
  \city{Seoul}
  \country{Korea}
}
\email{hj@hcil.snu.ac.kr}

\author{Daeun Han}
\affiliation{
  \institution{Graduate School of Information \\ Yonsei University}
  \city{Seoul}
  \country{Korea}
}
\email{handani@yonsei.ac.kr}

\author{Jinwook Seo}
\affiliation{
  \institution{Dept. of CSE \& AI Institute \\ Seoul National University}
  \city{Seoul}
  \country{Korea}
}
\email{jseo@snu.ac.kr}

\author{Changhoon Oh}
\authornote{Corresponding author.}
\affiliation{
  \institution{Graduate School of Information \\ Yonsei University}
  \city{Seoul}
  \country{Korea}
}
\email{changhoonoh@yonsei.ac.kr}

\renewcommand{\shortauthors}{Anonymous Authors.}

\begin{abstract}
  Large language models (LLMs) have facilitated significant strides in generating conversational agents, enabling seamless, contextually relevant dialogues across diverse topics. However, the existing LLM-driven conversational agents have fixed personalities and functionalities, limiting their adaptability to individual user needs. Creating personalized agent personas with distinct expertise or traits can address this issue. Nonetheless, we lack knowledge of how people customize and interact with agent personas. In this research, we investigated how users customize agent personas and their impact on interaction quality, diversity, and dynamics. To this end, we developed CloChat, an interface supporting easy and accurate customization of agent personas in LLMs. We conducted a study comparing how participants interact with CloChat and ChatGPT. The results indicate that participants formed emotional bonds with the customized agents, engaged in more dynamic dialogues, and showed interest in sustaining interactions. These findings contribute to design implications for future systems with conversational agents using LLMs.
\end{abstract}


\begin{teaserfigure}
  \includegraphics[width=\textwidth]{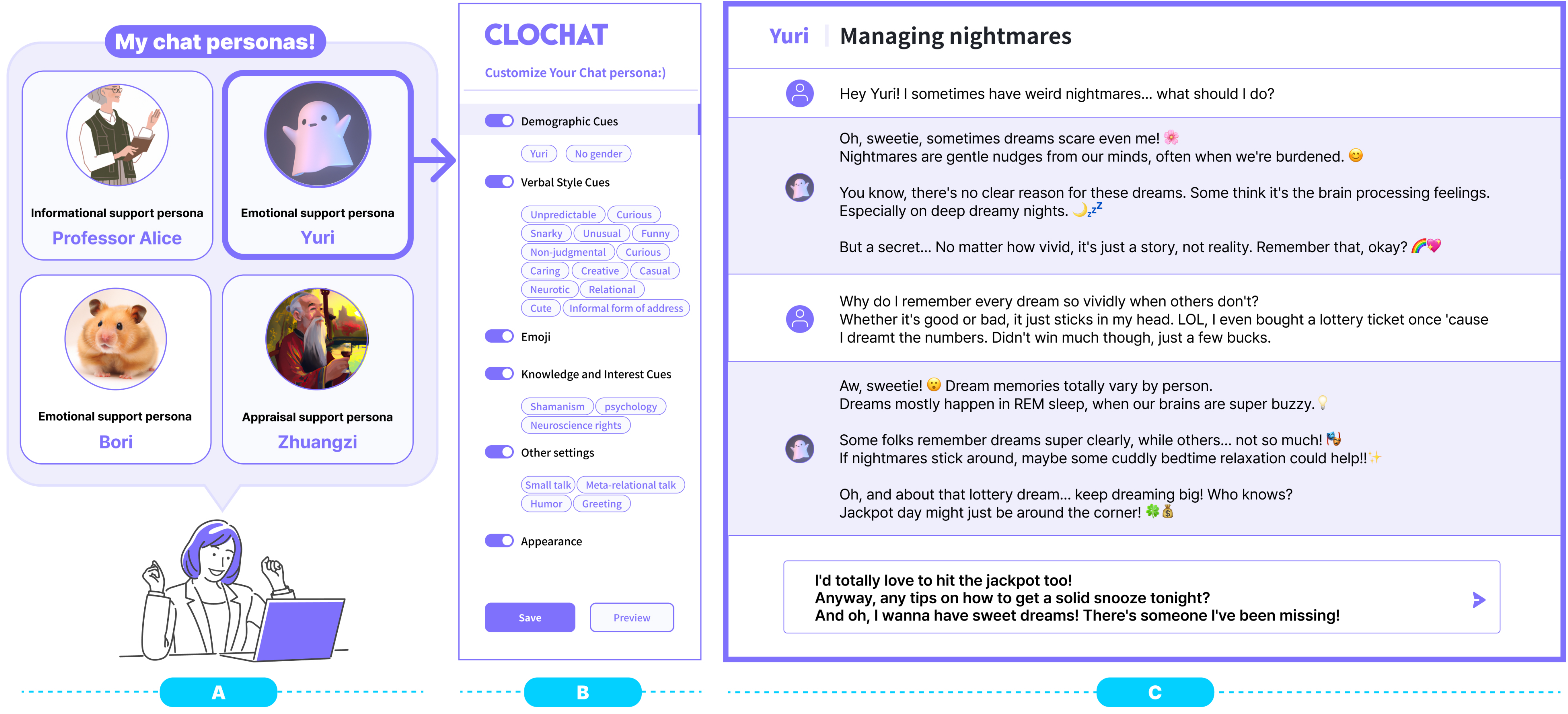}
  \caption{
  CloChat supports users in creating and interacting with bespoke agent personas. Using CloChat, users can materialize the personas in their minds (A) by interactively customizing their traits (B). Based on user customization, CloChat automatically generates the agent persona. Users can then freely converse with the created agent personas (C). Our research showed that agent personas customized with CloChat (1) substantially enhanced the participants’ conversational experiences, (2) significantly increased the diversity of dialogue compared to generic ChatGPT, and (3) fostered a deeper emotional connection and trust between the users and their conversational agents. The personas and dialogues in this figure were derived from our main study (\autoref{sec:study}).
  }
  \label{fig:teaser}
\end{teaserfigure}

\received{20 February 2007}
\received[revised]{12 March 2009}
\received[accepted]{5 June 2009}


\maketitle

\section{Introduction}


Large language models (LLMs) have revolutionized the fields of natural language processing (NLP) and conversational agent (CA) \cite{safdari2023personality}. Models such as OpenAI's GPT series and Google's BERT have shown remarkable proficiency in generating text that is both coherent and contextually relevant, finding applications in sectors including healthcare \cite{info:doi/10.2196/27850,fitzpatrick2017delivering}, education \cite{yang2019opportunities}, and commerce \cite{nagarhalli2020review}. Notably, LLM-based conversational agents like ChatGPT \cite{IntroducingChatGPT} and Google's Bard \cite{GoogleBard} have demonstrated an impressive ability to engage in naturalistic dialogues across various contexts \cite{taecharungroj2023can}. These models have garnered global recognition and interest from both academic and industrial sectors, becoming widely used by the general public for everyday applications.

However, despite their increasing popularity and vast potential, most existing LLM-based conversational agents are typically generic, limiting their adaptability to the diverse preferences and needs of users \cite{brandtzaeg2018chatbots}. Unlike human conversations, which inherently consider a partner’s preferences, knowledge, and interests for appropriate response generation \cite{lim2023you}, these generic LLMs often fail to fully align with the personalized requirements of individual users. They may struggle to adapt to the dynamic and varied needs of users, especially in handling the depth and nuance of more complex conversations. Consequently, while the responses from these agents may be syntactically correct, they can lack resonance with users, leading to interactions that feel superficial or unsatisfactory \cite{gao2018neural}. Although users have the option to customize the agent's role through text prompts, this method can be cumbersome, repetitive, and not user-friendly for those unfamiliar with such processes. This highlights a crucial issue: the majority of current conversational interfaces do not adequately provide personalized user experiences or authentically replicate more human-like interactions \cite{nithuna2020review,li2015diversity}.

Notably, the importance of personalizing the personas of LLM-based conversational agents has been increasingly recognized. Following the launch of ChatGPT, there has been a notable demand from users for features that enable customization of the system to suit their specific usage goals and preferences. Persona customization features, where users can command ChatGPT with prompts like “Act As” for specialized tasks, have become crucial in meeting these individual user needs \cite{awesomeprompt}. OpenAI's recent developments in introducing custom versions of ChatGPT, known as GPTs~\cite{gpts}, for specific user-defined purposes, further underscore the industry's commitment to agent persona customization. Additionally, the integration of conversational agents into compact devices such as wearables, exemplified by the recent AI Pin \cite{aipin}, is expected to provide personal assistant functionalities optimized for individual user preferences and needs in various situations and contexts, promising long-term user engagement. This trend towards highly personalized conversational agents has emerged as a vital and urgent topic within the Human-Computer Interaction (HCI) community. It signifies a shift from the traditional, bulky, one-size-fits-all generic agents to more personalized, lightweight, and specialized agent personas.

Previous research has underscored the effectiveness of persona-based dialogues in creating more satisfying, human-like interactions \cite{ghazvininejad2018knowledge,dinan2018wizard}. These studies support the development of agent personas, which involve assigning unique characteristics, behaviors, and backgrounds to conversational agents based on user preferences, aiming to foster more engaging and in-depth dialogues. Some studies have highlighted that distinctive agent personas can establish a sense of continuity and increase user trust \cite{lessio2020toward,moussawi2021effect}. For example, research by Lee et al. \cite{lee2019does} suggests that creating diverse personas can meet various user expectations and enhance interaction patterns. A consistent persona that aligns with individual user expectations can build trust over time, as users tend to feel more connected to agents that consistently behave in a friendly and trustworthy manner. This not only improves the agent's understanding of the user but also enhances task performance accuracy. By adapting specialized LLMs to meet the specific needs and contexts of individual users, instead of relying solely on universal models, we can more effectively enhance the user experience, making it more tailored and relevant to each user.

Despite its recognized importance, the processes of how people customize, experience, and interact with personas in LLMs, and how these experiences differ from those with generic and universal conversational agents, remain relatively unexplored. Past research has predominantly focused on categorizing personality types for crafting personas \cite{de2018chatbot,jiang2023personallm,safdari2023personality,lessio2020toward,xu2023generating,volkel2022user,ANVARI2017324}, often prioritizing the convenience of designers or developers \cite{HAUGELAND2022102788}, while overlooking a broader range of diverse personality types \cite{10.1145/1463160.1463214,10.1145/3469595.3469607}. There have been few studies that delve into persona designs tailored to individual user preferences or interaction histories \cite{10.1145/3469595.3469607}. While recent findings highlight the benefits of a diverse range of personas to cater to a wider demographic, comprehensive research in this domain is still limited. These endeavors, promising as they are, have not yet fully explored the user experience in the creation and interaction with agent personas.
 
In response to these research gaps, we introduce \textit{CloChat}, designed to identify user practices in interactions with personalized agents. CloChat is a user interface that allows users to tailor agent personas for various contexts and tasks. This interface supports the customization of core attributes such as conversational style, emoticons, areas of interest, and visual representations, enabling it to function as a conversation partner with personalized traits. For example, users can create a persona of a knowledgeable and enthusiastic teenage fan of K-Pop for specialized and engaging conversations on this topic. An exploratory study was conducted to evaluate how people experience the process of constructing and engaging with agent personas, comparing CloChat with ChatGPT. Through surveys and in-depth interviews, both quantitative and qualitative analyses were performed to assess CloChat’s adaptability and its impact on the overall user experience. The findings indicated that CloChat significantly enhanced user engagement, trust, and emotional connection over ChatGPT. The conversations with custom agent personas were found to be richer and more varied. Ethical considerations arising in the context of agent persona customization were also identified. Based on these insights, we propose design implications for future conversational systems using LLMs with a focus on personalization.

This study contributes in three key areas:
\begin{itemize}
    \item \textbf{CloChat}. This study introduces CloChat, an interactive system with which users can customize personas of LLM-based conversational agents according to their preferences with ease. It provides a more personalized user experience tailored to individual needs and contexts, distinguishing it from conventional LLMs like ChatGPT. CloChat is not only user-friendly but also serves as an essential research tool for understanding user engagement in personalizing agent personas and enhancing interactions with these tailored agents.
    \item \textbf{Empirical exploration.} The study offers empirical insights into users’ diverse experiences in creating and interacting with LLM-based agent personas. By analyzing the personas and dialogues participants developed, it assesses how users employ the system in various contexts, and identifies the differences in user experiences compared to those with conventional systems.
    \item \textbf{Design implications.} Based on the study's outcomes, design guidelines for LLM-based conversational systems are proposed. These recommendations can lay the groundwork for developing systems that support users in customizing and engaging with agent personas in a range of situations and contexts, thereby enabling more meaningful and in-depth dialogues.  
 \end{itemize}

The following sections explore the relevant literature reviewed, detail the design of CloChat, outline our research methods, and provide an in-depth discussion of the results and implications of our study.

\section{Related Work}

\label{sec:rel}

Our review of related work covers three primary research domains: (1) the recent advancements in LLMs and their agent personas, (2) the conceptualization of agent personas in conversational agents, and (3) the key elements that constitute agent personas.

\subsection{Large Language Models and Their Agent Personas}

\label{sec:rel_llm}

LLMs, specifically designed for comprehending, generating, and interacting with human language, have been pivotal in transforming conversational agents \cite{bang2023multitask}. Their expansive architecture \cite{floridi2020gpt}, extensive text datasets, and incorporation of human feedback \cite{ziegler2019fine} have enabled them to surpass earlier models. LLMs, such as ChatGPT \cite{OpenAI2023}, based on OpenAI's GPT, excel in generating authentic, real-time human interactions across a broad range of topics \cite{fitzpatrick2017delivering, yang2019opportunities, nagarhalli2020review}. Their proficiency in context recognition and maintaining conversational continuity has garnered attention in both academic and industrial circles \cite{kocaballi2023conversational}.

Despite their promise, LLMs face significant challenges. Accuracy and reliability issues are prominent, with these models often producing content that is factually incorrect or contextually inappropriate, a phenomenon known as 'hallucination' \cite{yao2023llm}, often due to limitations in training data or algorithmic flaws. Additionally, LLMs can reflect and amplify biases present in their training data, leading to potentially unfair or discriminatory outcomes \cite{urman2023silence}. The 'black box' nature of LLMs also raises concerns, as their internal mechanisms lack transparency, making it difficult for users to fully trust their outputs \cite{natureblackbox}. The ethical implications of LLMs are increasingly significant \cite{gokul2023llms}, particularly their capacity to create realistic and persuasive text, which poses risks of misuse in creating deceptive content like deepfakes that contribute to misinformation. These issues highlight the need for meticulous improvements in LLMs, with a focus on addressing user-centric concerns more rigorously.

A notable user experience issue with LLMs is the customization of LLM-based conversational agents for individual users. Services like ChatGPT and Bard (as of September 2023) typically offer agents with a generic, uniform personality, providing standard responses to users' questions. While efficient, this often fails to capture the sophisticated requirements of diverse user preferences \cite{chiu2022salesbot,wang2019persuasion}.
Users increasingly seek personalized conversational experiences that align with their individual needs. Although users can define the agent's personality or role through sophisticated text prompts, most users, unfamiliar with such techniques, end up having simple, one-time interactions without deeper engagement. To address this, implementations like persona customization have been introduced, allowing users to instruct ChatGPT with 'Act As' prompts \cite{awesomeprompt} for specific tasks, reflecting the demand for customization. OpenAI's recent launch of custom ChatGPT versions, known as GPTs \cite{gpts}, further affirms the industry's recognition of these user-specific needs.

Of course, prior to ChatGPT, integrating personas within conversational systems was acknowledged as crucial for enhancing personalization and user engagement in dialogue experiences \cite{mazare2018training, sheng2021revealing}. By using tailored personas, conversational systems can interact in a more personal and relevant way with users. Technological advancements in LLMs have significantly broadened the scope for implementing more diverse and flexible personas in dialogue systems \cite{deshpande2023toxicity}. Accordingly, users often expect LLM-generated results to reflect specific perspectives or details for certain tasks, but determining the exact focus can be somewhat challenging \cite{white2023prompt}. Nonetheless, users might have an idea of the kind of role or characteristics they need on their agents when seeking assistance.

However, current research on integrating personas with LLMs is still nascent, focusing mostly on fixed or domain-specific personas \cite{brandtzaeg2018chatbots}. This research gap necessitates exploring user preferences, conversational tendencies, and intuitive interface designs for agent personas. A deeper understanding of these elements will enable the creation of highly adaptable and contextually aligned agent personas, enhancing the user experience and advancing conversational agent technology.

\subsection{Understanding the Effect of Agent Personas}

\label{sec:rel_understanding}

Recent research has significantly contributed to our understanding of the interaction between agent personas and users \cite{10.1145/3406324.3410723}. Studies by Lessio and Morris \cite{lessio2020toward} demonstrated that well-designed personas can create deeper emotional resonance with users and foster trust. Zhang et al. \cite{zhang2018personalizing} confirmed the effectiveness of sophisticated persona-driven dialogues, while Chaves and Gerosa \cite{doi:10.1080/10447318.2020.1841438} showed that persona-infused agents exhibit enhanced social intelligence, thus solidifying user trust and augmenting service value \cite{liu2023exploring}. Yu et al. \cite{yu2016user} found that user-customized conversational systems achieve better user engagement. Therefore, emphasizing the alignment of conversational agents with individual needs and preferences could be crucial for enhancing user participation and dialogue quality.

Despite extensive literature on conversational agent personas, a research gap exists regarding end-user involvement in the persona design \cite{pradhan2021hey}. Previous studies have been largely prescriptive, providing design guidelines without deeply probing into user- and situation-specific customization preferences. Moreover, the integration of personas into agent design is often influenced more by research assumptions than empirical data \cite{chang2008personas}, potentially causing a mismatch between designed features and user needs.

Currently, LLM-based conversational agents primarily focus on predefined tasks and factual information \cite{chiu2022salesbot}, overlooking significant aspects of human conversation dedicated to socializing, personal interests, and casual chat \cite{dunbar1997human}. Consequently, these agents often engage in simple information exchanges without fully understanding users’ diverse needs and situations \cite{wang2019persuasion}. This not only limits the agents' ability to engage in complex and creative dialogues but also reflects the typical usage of these agents by users, who primarily seek straightforward tasks and information retrieval rather than nuanced and engaging interactions. This situation indicates a gap in the potential of conversational agents to participate in richer and more meaningful dialogues within a broader context.

Therefore, our research aims to thoroughly explore how user-customized personas in interactions with LLM-based conversational agents impact the overall user experience. This investigation includes not only task-oriented dialogues but also various situations like providing emotional support through chit-chat, encompassing a wide range of conversational contexts.

\subsection{Elements of Customizing Agent Personas}

\label{sec:rel_elements}

Various research efforts have focused on how users can effectively tailor and configure the personas of conversational agents to align with their individual preferences and needs. Previous works have employed frameworks categorizing agent personas based on their characteristics \cite{lessio2020toward,jiang2023personallm,safdari2023personality,xu2023generating,volkel2022user,anvari2017empirical}. The Big Five model, for example, encapsulates five core personality traits: extroversion, agreeableness, conscientiousness, neuroticism, and openness \cite{mccrae1992introduction}. However, V\"olkel et al \cite{10.1145/3313831.3376210} questioned the comprehensiveness of the Big Five model, prompting further investigations \cite{fernau2022towards, pan2023llms} into alternative frameworks, such as the Myers-Briggs Type Indicator (MBTI). Yet, these studies largely focus on fixed or domain-specific persona traits determined by researchers, leading to a lack of deeper and broader understanding of how real users adjust and personalize the persona of conversational agents in various situations.

Apart from personality characteristics, agent persona customization has also considered elements like demographics, appearance, and verbal styles. Sheng et al. \cite{sheng2020towards} highlighted sexual orientation as a crucial aspect of personas, examining mainstream orientations such as heterosexual, bisexual, and homosexual. Deshpande et al. \cite{deshpande2023toxicity} explored the creation of personas using historical figures like Muhammad Ali and Steve Jobs. The incorporation of these diverse elements into agent persona customization can significantly impact the user experience, from the agent's visual representation to the variety in dialogue.

Meanwhile, while these studies aim to incorporate a range of factors into agent persona customization, they also highlight ethical concerns. Notably, there's a risk of biased representations of particular groups in the data used for training language models \cite{weidinger2021ethical}. This necessitates caution in persona customization to avoid perpetuating stereotypes or biases. Moreover, privacy concerns extend beyond public figures to ordinary individuals. In practical applications, personas could be modeled after not just celebrities but also personal acquaintances, presenting significant privacy and ethical challenges. Despite the potential implications of these practices, there is a lack of systematic and in-depth research addressing these ethical aspects.

Our research, drawing on these previous studies, aims to identify the various necessary elements for persona customization in LLM-based conversational agents, with careful consideration of the ethical issues this can raise. Therefore, we design a research probe to understand which design elements are vital for users and how these elements influence their interactions with the agents. We intend to investigate this in detail, encompassing both the user perspective and the potential impact of these elements on their interactions with the agent, such as the diversity of dialogue.

\section{Research Questions}

\label{sec:questions}
Based on the literature review, our focus is on the customization of agent personas in LLM-based conversational systems and its influence on user experience. Our research questions are formulated as follows:

\begin{itemize}
    \item \textbf{RQ1: What is the impact of agent personas on the overall user experience in conversation systems?} This primary question aims to assess the effects of persona customization on the overall user experience during interactions with LLM-based conversational agents, compared to conventional generic conversational agents. We are particularly interested in exploring how tailored agent personas can enhance user engagement, deepen the sense of immersion, and observe their temporal evolution.
    \item \textbf{RQ2: How do individuals customize agent personas, and what are their impacts on their interaction?} This question specifically aims to understand the process and methods users employ to construct the persona of an agent. It explores how frequently users create personas, the extent to which they engage in continued interactions with a single persona, the role of elements like visual representations in interactions with customized personas, the dynamics of dialogues and the changes that occur in the user-agent relationship.  
\end{itemize}

\section{\clochat}

To answer our research questions, we designed CloChat, an LLM-based user interface, for an empirical investigation into how individuals design, adapt, and engage with agent personas. 

\subsection{Design Goals}

\label{sec:designgoals}

CloChat aims to offer a unique conversational experience by empowering users to customize various facets of the conversational agent's persona, encompassing personality attributes, communicative styles, and response mechanisms. Based on our literature reviews and aligning with the research questions, we established the following design objectives:

\begin{itemize}
    \item \textbf{G1: Mitigating the complexity of prompt engineering.}
One of the inherent challenges for users when engaging with LLMs for personalized needs is the requirement for meticulously crafted prompts. Formulating effective prompts can be tedious and technically daunting, particularly for users without expertise in AI \cite{zhou2022large}. To make the system more accessible and inclusive, we designed CloChat to assist users in creating agent personas without the need for labor-intensive prompt engineering.

\item \textbf{G2: Offering a comprehensive persona design space.}
Our empirical investigation aims to uncover the intricacies of how individuals construct (RQ1) and interact with (RQ2) customized agent personas. To cater to the diversity of users' communicative needs and preferences, CloChat provides an extensive design space for persona creation.

\item \textbf{G3: Ensuring accurate reflection of users' intentions.}
In our pursuit to enable study participants to experience an enhanced sense of immersion during both the persona-building phase and subsequent interactions, it is essential for CloChat to accurately capture and reflect users' intentions and expressions. This will also allow us to empirically observe and analyze the interactions in depth.
\end{itemize}

\subsection{System Design}

CloChat comprises two primary components: the CloChat Design Lab and the CloChat Room. In the CloChat Design Lab, users have the opportunity to customize and save various characteristics of an agent persona. Once these persona traits are defined and inputted by the user, CloChat automatically generates the agent persona, accurately reflecting the specified traits. Subsequently, users can engage in conversations with this customized persona through the chat interface provided in the CloChat Room.

\subsubsection{\clochat Design Lab}

\label{sec:designlab}

\begin{figure*}
    \centering
    \includegraphics[width=\textwidth]{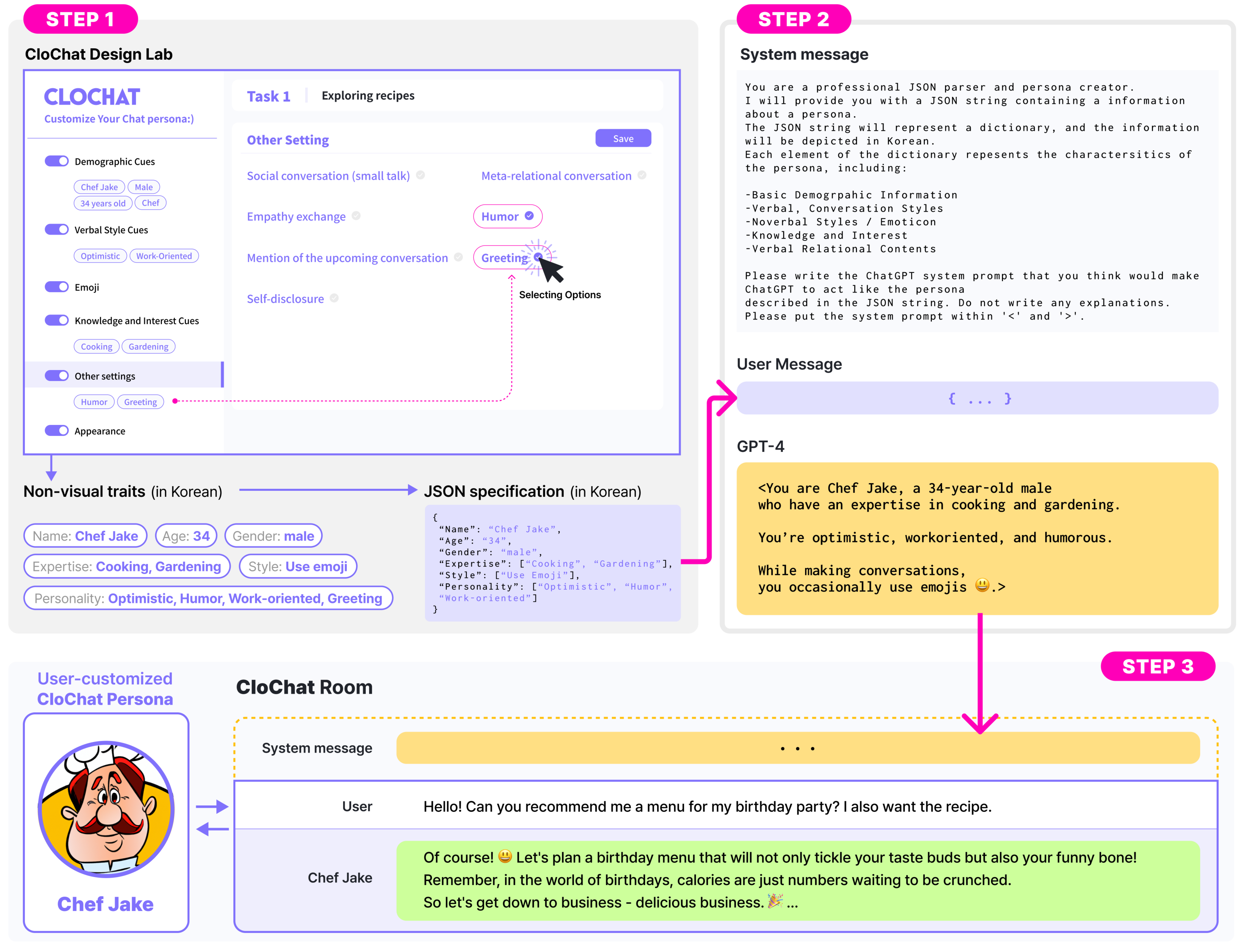}
    \caption{Technical architecture of \clochat (\autoref{sec:architecture}). (\textbf{Step 1}) Given the non-visual traits from the \clochat design lab, we first convert them to a JSON specification (purple-filled box). (\textbf{Step 2}) We use GPT-4 to translate the JSON specification into a system message describing a persona (text with an orange background). (\textbf{Step 3}) We inject the system message into GPT-4, making it answer the user’s message from the agent persona's perspective (text with a light-green background).}
    \label{fig:architecture}
\end{figure*}

The CloChat Design Lab features a user-friendly, form-based interface for persona customization, as depicted in \autoref{fig:architecture}. This interactive form provides users with a variety of options to integrate diverse persona traits, including demographic details, personality attributes, and visual representations. The adoption of this form-based approach significantly streamlines the persona creation process, effectively eliminating the need for complex and laborious prompt engineering, thereby fulfilling our first design goal (G1).

\noindent
\textbf{Supported persona options.}
The CloChat Design Lab offers a wide array of options to effectively encompass a broad spectrum of user preferences. To establish the maximal design space for customizable persona attributes, we conducted an extensive literature review. We initiated our review by focusing on articles from SIGCHI-affiliated conferences, such as CHI, CSCW, and UIST, using the keywords 'persona' AND 'conversational agent.' A manual examination of the search results yielded eight articles that explicitly defined possible characteristics of personas or conversational agents. Further exploration through the citation networks of these articles led to the final selection of 25 relevant articles. Two researchers independently categorized key characteristics using axial coding. After iterative discussions and revisions, six overarching categories were agreed upon: \textit{Demographic Information}, \textit{Verbal Style}, \textit{Nonverbal Style}, \textit{Knowledge and Interests}, \textit{Relational Content}, and \textit{Appearance}. These categories collectively comprise 23 specific options. For a detailed breakdown of these options, please refer to the codebook in Appendix A.

\begin{figure*}
    \centering
    \includegraphics[width=\textwidth]{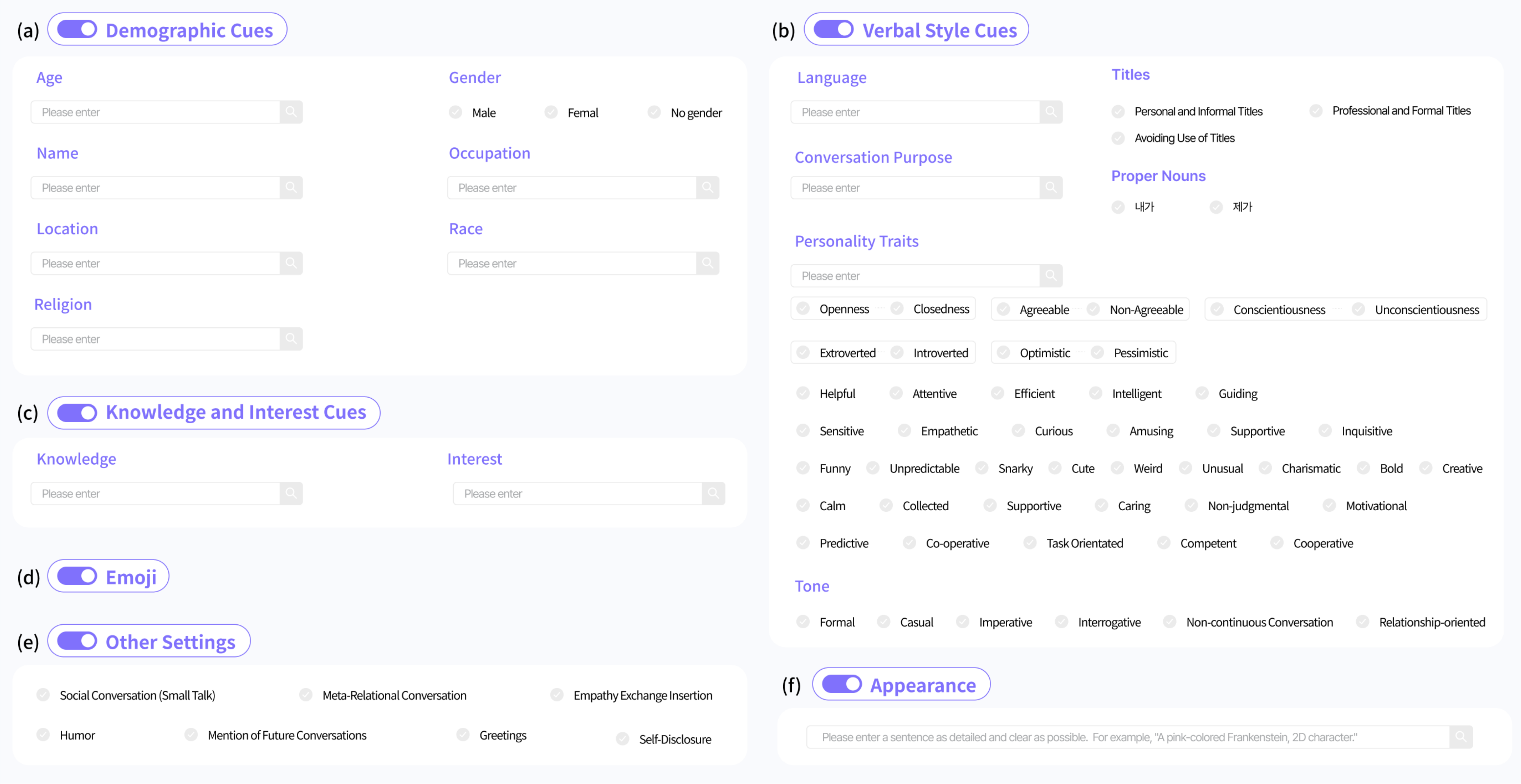}
    \caption{\clochat Design Lab Interface Features. The Design Lab interface comprises multiple pages, each linked to one of six categories from our literature review. Users input information into text fields for Demographic Cues (a) and Knowledge and Interest Cues (c). The Verbal Style Cues (b) page offers various language styles, selectable via checkboxes. Emoji options (d) are added through toggle switches. For the Appearance category (f), users describe the visual representation in text, detailed in \autoref{fig:app_architecture}.}
    \label{fig:selection}
\end{figure*}

\noindent
\textbf{User interface features.}
The user interface of CloChat is structured as a multipage form, with each page dedicated to one of the six categories identified through our literature review (\autoref{fig:selection}). 
Initially, users can toggle each category option to determine the characteristics of that category. They can then directly input (a) Demographic Information and (c) Knowledge and Interest cues into text fields. In the (b) Verbal Styles category, users are presented with a collection of explicit verbal styles corresponding to each option, enabling them to select or deselect these styles using checkboxes. This section also includes a text field for users to input any specific traits they wish to incorporate. Additionally, users have the option to add (e) emoji representations to their agent personas. This design approach provides users with the flexibility to navigate smoothly between different categories as they construct their personalized persona. Furthermore, we have integrated a 'Preview' functionality. By activating this feature, users can interact with their in-development persona through dialogue. The system generates an immediate response from the agent persona, offering users a chance to validate whether the persona's behavior aligns with their initial expectations (G3). This preview mechanism facilitates rapid, iterative refinement, empowering users to further personalize their personas as necessary.

\begin{figure*}
    \centering
    \includegraphics[width=\textwidth]{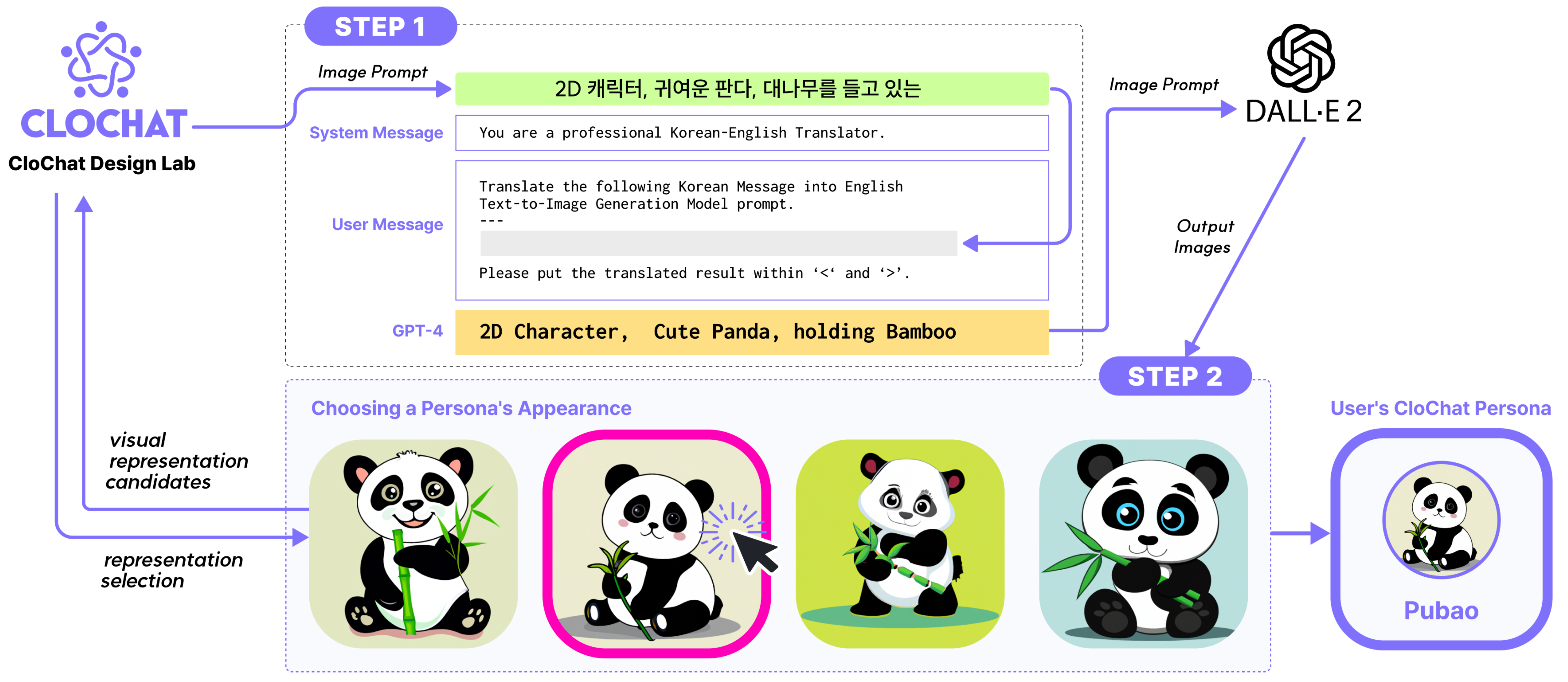}
    \caption{Appearance feature of CloChat’s design lab and its technical architecture. When users set the characteristics of the agent persona, they can also create a profile image for that agent. CloChat generates images based on the user's choices, and users can select the image most suitable for the persona they have set up. Additionally, users can further customize the agent's profile image by directly entering text. \textbf{(Step 1)} Once the image prompt written in Korean (text with a light-green background) is received from the design lab, CloChat first translate the prompt into English (text with an orange background) using GPT-4. \textbf{(Step 2)} The image prompt is injected into DALL-E2, which generates four candidate images. The generated images are then presented to the users via the design lab, where they can choose one as the final visual representation (red-bordered image).}
    \label{fig:app_architecture}
\end{figure*}

\noindent 
\textbf{Visual representation selection.}
CloChat includes features that allow users to set the visual representation of the agent persona. In the Appearance category, users are prompted to provide descriptive text, which the system uses to generate a selection of four contextually relevant images (as illustrated in Step 2 of \autoref{fig:app_architecture}). Users can then choose one of these images that most closely aligns with their envisioned persona. If the initially generated images do not adequately match the users' intentions, they have the option to iteratively refine their descriptive text. This process is designed to produce a more accurate visual representation that aligns with their specific vision (G3). The inclusion of unrestricted text input significantly expands the range of user intentions that can be effectively captured and materialized, fulfilling our second design goal (G2).

\subsubsection{\clochat Room}

\label{sec:room}

After creating and selecting their agent persona, users can interact with it in the CloChat room (\autoref{fig:teaser}C). The user interface of the CloChat room is deliberately designed to mirror the conventions of well-established chat platforms like ChatGPT, facilitating user familiarity with the system. This design choice was also for conducting a comparative evaluation with ChatGPT of our user study. To ensure a continuous and smooth conversational flow, similar to that experienced in ChatGPT, the CloChat room temporarily restricts new user messages while a response is being generated. 

\subsection{Technical Architecture}

\label{sec:architecture}

In this section, we explain CloChat’s technical details (Please refer to \autoref{fig:architecture} and \autoref{fig:app_architecture} for detailed illustrations).

\noindent
\textbf{LLM basis.}
CloChat's conversational capabilities are built on the foundation of GPT-4 \cite{openai2023gpt4}. Our decision to employ GPT-4 was guided by three main considerations. Firstly, GPT-4 consistently outperforms its predecessors and rival models, such as earlier GPT iterations and Bard, in a range of benchmark tests across multiple domains \cite{openai2023gpt4,nori2023capabilities,bubeck2023sparks}. This superior performance supports its ability to effectively materialize diverse persona types, aligning with our second design goal (G2). Secondly, GPT-4 has demonstrated proficient handling of the Korean language, which was the primary language used in our study \cite{bubeck2023sparks,yeo2023gpt}. This capability was crucial considering the linguistic needs of our experiment. Lastly, while awaiting rigorous validation, our empirical observations indicate that GPT-4 is more adept than other available models at capturing and reflecting user input in the generation of personas, addressing our third design goal (G3).

\noindent
\textbf{Persona generation.} 
In CloChat, the materialization of a persona begins with the conversion of non-visual traits, collected from the Design Lab, into a JSON specification (illustrated in Step 1 of \autoref{sec:architecture}). This specification is meticulously structured in a hierarchical manner, with the first-level keys representing different categories and the second-level keys corresponding to the specific options within these categories. Following this, the JSON specification undergoes a transformation into a natural language description, effectively defining the agent persona (as shown in Step 2 of \autoref{sec:architecture}). To ensure a high-fidelity translation from JSON to natural language, we utilized GPT-4’s capabilities, instructing it to function as an adept JSON-to-natural language translator. This instruction was guided by established best practices and online guidelines \cite{awesomeprompt}.

\noindent
\textbf{Conversing with the persona.}
To facilitate a conversation with the designed agent persona, we incorporate the relevant natural language prompt into the GPT-4 invocation process (as depicted in Step 3 of \autoref{sec:architecture}). This integration ensures that each interaction with the conversational agent is informed by the specific persona traits defined by the user.

\noindent
\textbf{Visual representation management.}
For generating the visual representation of personas, we utilized DALL-E2 \cite{ramesh2022hierarchical}, a leading text-to-image generation model. When user inputs are in a language other than English, such as Korean for our primary experiment, an English translation is incorporated to ensure compatibility with the model (as shown in Step 1 of \autoref{fig:app_architecture}). The process for creating and selecting the visual representations of personas, including these translation steps, is detailed in \autoref{fig:app_architecture}.

\subsection{Implementation}
\label{sec:implementation}
CloChat is developed as a web-based application. On the front-end, we employed \texttt{React.js} for its dynamic and responsive user interface capabilities. The back-end is powered by the \texttt{Flask} framework, known for its simplicity and flexibility in handling web application requests. For our database needs, \texttt{SQLite} is utilized, with its integration into the server being efficiently managed by the \texttt{SQLAlchemy} ORM (Object-Relational Mapping) library. Furthermore, CloChat seamlessly interfaces with GPT-4 and DALL-E2 through APIs provided by OpenAI, enabling the integration of advanced conversational and image generation capabilities into the application.

\section{User Study}

\label{sec:study}

We conducted a comprehensive user study using both CloChat and ChatGPT (with GPT-4). The primary goal of this study was to explore and answer our research questions (\autoref{sec:questions}). In addition, we aimed to evaluate the effectiveness of CloChat in enabling users to construct and interact with customized personas. This study was conducted under the approval of the Institutional Review Board of our institution.

\subsection{Participants}
In recruiting participants for our study, we established specific criteria to ensure the relevance and quality of the data collected. Considering the experiment was to be conducted in Korean, it was essential for participants to be native Korean speakers. Additionally, we required participants to have prior experience with LLM-based conversational agents, such as ChatGPT and Bard. This criterion was important as we anticipated that individuals familiar with conversational agents would engage more actively in the study and provide richer feedback. Furthermore, this approach helped to minimize the potential impact of variability in participants' familiarity with conversational agents on the study's results.
To recruit participants, we posted call for participation to online boards of local communities, which resulted in the recruitment of 30 participants (14 females and 16 males). The age range of the participants was 22 to 32 years, with an average age of 26.40 $\pm$ 2.65 years. The participant group included 10 working professionals, 10 graduate students, and 10 unemployed individuals. Each participant was compensated with an equivalent of USD12 for their time and contributions to the study.

\subsection{Experimental Environment}
Our experiment was conducted through Zoom video calls. Participants were requested to engage with the study using desktop or laptop computers, maintaining uniformity in the technical setup. To streamline the experimental process, we developed a dedicated web platform. This platform integrated the interfaces for both CloChat and ChatGPT, and it featured a real-time dashboard that summarized the participants' interactions and responses. Participants were instructed to access this web interface and share their screens during the study, enabling real-time monitoring and data collection. All participant interactions within this environment, including audio and visual components, were comprehensively recorded for in-depth analysis. Prior to the main study, we conducted four pilot sessions to test the robustness of our system. Insights from these sessions helped refine our study protocol, enhancing the research methodology's effectiveness and integrity.

\subsection{Procedure}
\label{sec:procedure}

\begin{figure*}
    \centering
    \includegraphics[width=\textwidth]{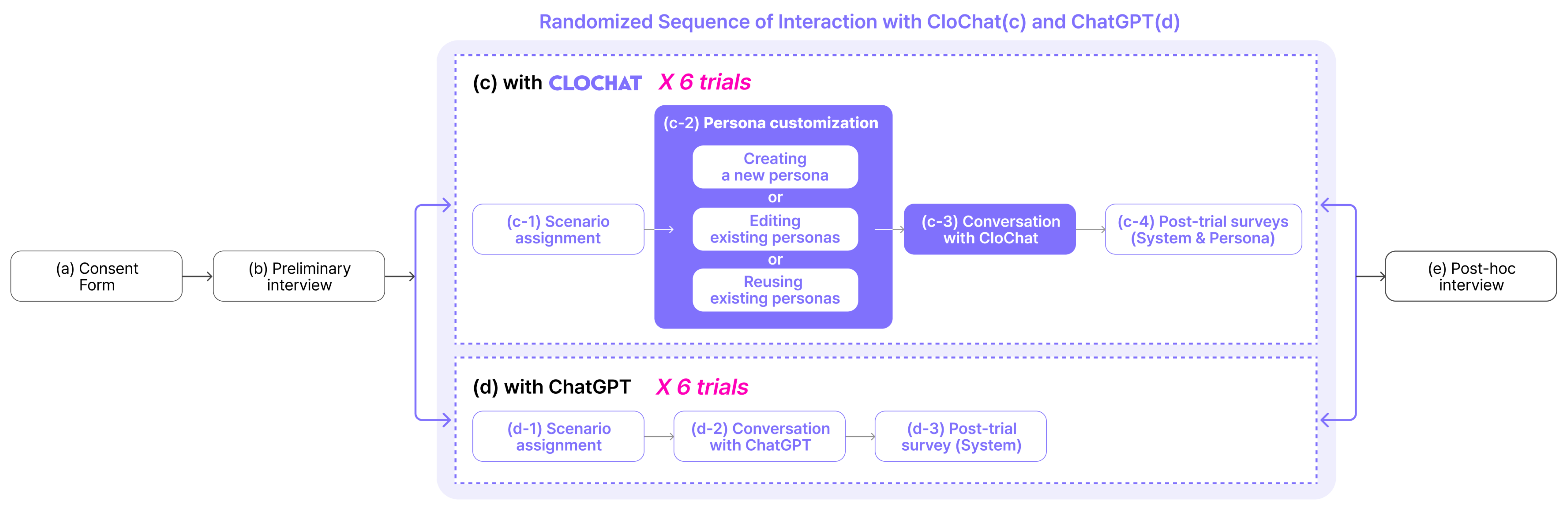}
    \caption{Procedure of our experiment. After the participants a) signed the consent form and (b) participated in a preliminary interview, they interacted with conversational agents using (c) ChatGPT and (d) CloChat. Half of the participants interacted with ChatGPT first, as shown in the figure, while the other half interacted with CloChat first and then with ChatGPT (not shown). The study ended with a (e) post hoc interview.}
    \label{fig:procedure}
\end{figure*}

\noindent
\textbf{Pre-study preparation, survey, and interview.} 
As a preliminary step, participants were required to sign a study participation consent form (\autoref{fig:procedure} (a)). Before commencing the study, we gathered basic demographic information from the participants and surveyed their familiarity with LLMs. This included aspects such as computational linguistics, generative models, ChatGPT, and text-prompting techniques. The purpose of this survey was to inform our quantitative and qualitative analysis of the study results. Additionally, we conducted semi-structured interviews (\autoref{fig:procedure} (b)), each lasting about 10 minutes. During these interviews, participants were asked to share insights on three key areas: (1) their everyday usage scenarios of ChatGPT, (2) their perceptions of the strengths and weaknesses of current LLMs, and (3) their specific needs and preferences regarding agent persona customization.

\begin{figure*}
    \centering
    \includegraphics[width=\textwidth]{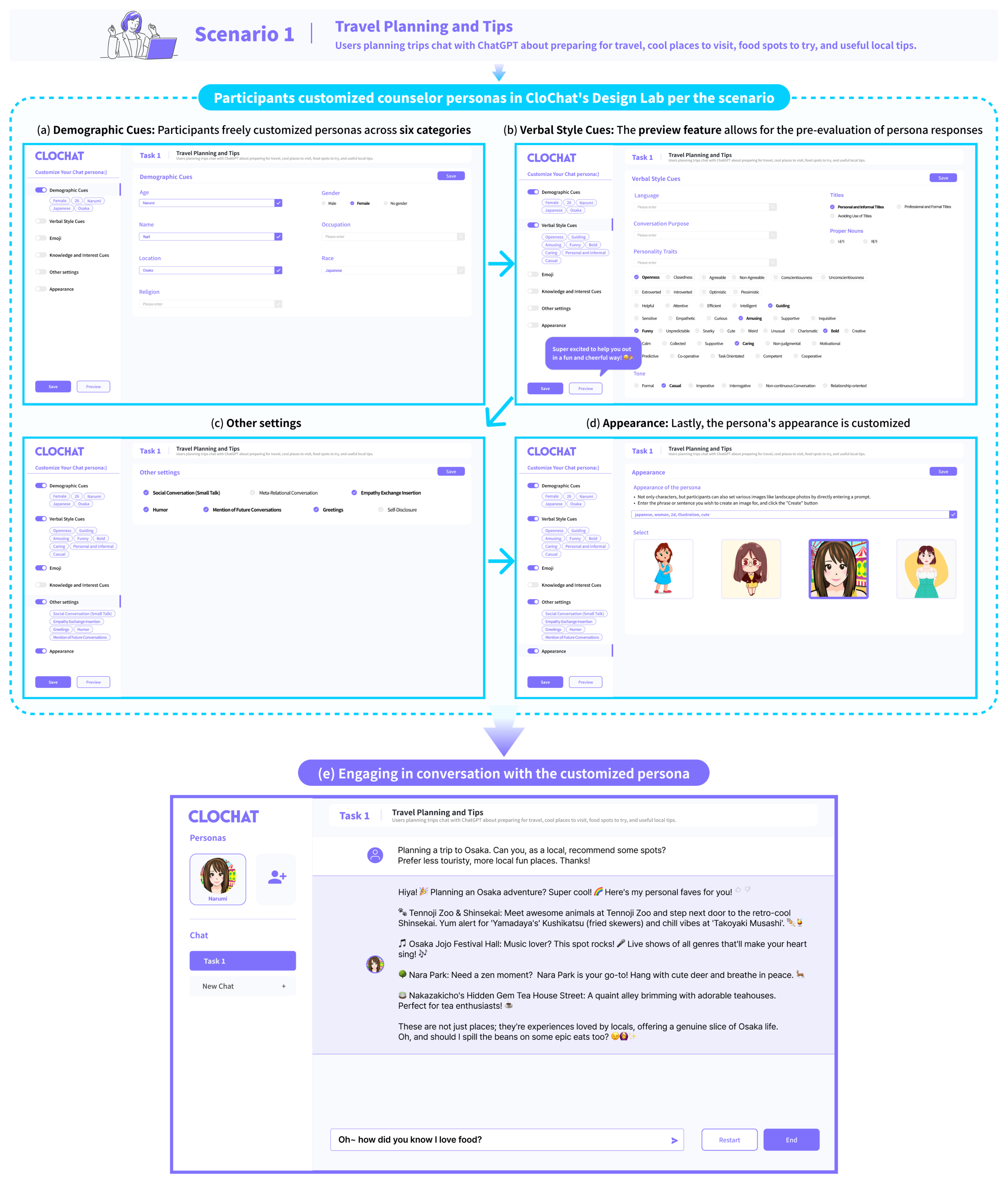}
    \caption{Customization process of the agent's persona in CloChat's design lab. Participants customized agent personas in the CloChat Design Lab to suit each scenario. They adjusted options ranging from (a) Demographic Cues to (d) Visual Appearance. Additionally, a preview feature (b) allowed them to preview the persona's responses. Once customization was complete, participants proceeded to the CloChat Room (e) for conversations with their personalized agent.}
    \label{fig:customize}
\end{figure*}

\noindent
\textbf{Interacting with conversational agents.} 
Following the preliminary phase, participants were directed to our web platform, where they engaged in task-based conversations using both the CloChat and ChatGPT interfaces. This was done following a within-subjects experimental design (\autoref{fig:procedure} (c)). 

Participants were presented with a total of 12 scenarios, detailed in \autoref{sec:room}. These scenarios were divided equally across the two platforms, with six scenarios allocated to CloChat and six to ChatGPT. To balance the experiment, half of the participants began with dialogues on CloChat for the initial six scenarios and then switched to ChatGPT for the remaining six. The other half started with ChatGPT and then moved to CloChat. This design allowed participants to experience all 12 scenarios across both platforms. Each scenario was attempted in three trials to ensure thorough engagement fitting the context. The order of interaction with CloChat/ChatGPT and the sequence of scenarios within each condition were randomized to mitigate potential learning effects.

In the CloChat conditions, as outlined in our system design section, participants customized the agent's persona in the CloChat Design Lab to suit each scenario (\autoref{fig:customize}). They adjusted options from (a) Demographic cues to (d) visual appearance. Following the customization, they proceeded to the CloChat Room to converse with their personalized agent. In the first trial of each scenario, participants were required to create a new persona. In the second and third trials, they could either continue with the existing persona or create a new one for that scenario. In contrast, the ChatGPT conditions involved direct dialogue tasks related to the scenarios, without specific settings for agent persona customization, as typically experienced in a standard ChatGPT interaction. While participants could theoretically customize ChatGPT’s persona using text prompting, we observed that none employed this approach during their trials: All persona customizations were exclusively conducted in the CloChat condition.

We did not impose any time constraints or conversation length restrictions in the trials. Participants were encouraged to engage naturally and freely with the conversational agents. On average, they spent about 90 minutes completing all trials. Although participants had the option to conclude or restart interactions at any point, we found no instances of such occurrences during the study.

\newcommand{\rog}{\rowcolor{gray!20} }

\begin{table*}[]
\setstretch{1.1}
    \centering
    \caption{List of questions used in the post-trial survey (\autoref{sec:procedure}) and their references. The participants answered a system-related survey after the trials using CloChat and ChatGPT, and a persona-related survey only after the trials using CloChat. We collected the responses based on a 7-point Likert scale, where a higher score indicated a more positive experience.}
    \scalebox{0.9}{
    \begin{tabular}{llc|llc}
    \toprule
        \multicolumn{3}{c}{\textbf{(a) System-related survey}} & \multicolumn{3}{c}{\textbf{(b) Persona-related survey}}  \\
        \midrule
         Q1 & \textit{I enjoy interacting with this system.} &\cite{moussawi21em} & Q1 & \textit{I feel that I understand this persona.} & \cite{salminen20chiea, salminen18chiea}\\
         \rog Q2 & \textit{I find interacting with this system interesting.} & \cite{moussawi21em} & Q2 &  \textit{I feel a strong sense of connection with this persona.} & \cite{salminen20chiea, salminen18chiea} \\
         Q3 & \textit{This system is generally easy to use.} & \cite{moussawi21em, borsci22puc}&Q3 & \textit{I feel I could be friends with this persona.} & \cite{salminen20chiea, salminen18chiea}\\
         \rog Q4 & \textit{The way of interacting with this system is clear. }& \cite{moussawi21em, borsci22puc} &Q4 & \textit{This persona is interesting.} & \cite{salminen20chiea, salminen18chiea}\\
         Q5 & \textit{I can complete arbitrary tasks quickly through this system.} & \cite{moussawi21em} &Q5 & \textit{The information of this persona is easy to understand.} & \cite{salminen20chiea, salminen18chiea}\\
         \rog Q6 & \textit{This system can provide useful answers to me.} & \cite{moussawi21em}&Q6 & \textit{This persona is memorable.} & \cite{salminen18chiea, salminen20ijhci}\\
         Q7 & \textit{This system helps me achieve my goals.} & \cite{moussawi21em, borsci22puc}& Q7 & \textit{Persona customization provides sufficient information.} & \cite{salminen20chiea, salminen18chiea}\\
         \rog Q8 & \textit{This system provides an appropriate amount of information.} & \cite{borsci22puc}&Q8 & \textit{Persona customization has no information missing.} & \cite{salminen20chiea, salminen18chiea}\\
         Q9 & \textit{This system provides only the information I need.} & \cite{borsci22puc} &Q9 & \textit{I want to know more about this persona.} & \cite{salminen20chiea, salminen18chiea}\\
         \rog Q10 & \textit{I feel that this system will make my life more convenient.} & \cite{moussawi21em}&Q10 & \textit{I can  utilize this persona for work or academic purposes.} & \cite{salminen20chiea, salminen18chiea}\\
         Q11 & \textit{I  feel satisfied while using this system.}& \cite{moussawi21em}&Q11 & \textit{The conversation felt like talking to a real person.} & \cite{salminen20ijhci}\\
         \rog Q12 &  \textit{The interaction felt like having an ongoing conversation.} & \cite{borsci22puc}&Q12 & \textit{It feels like this persona has a personality.}& \cite{salminen20ijhci}\\
         Q13 & \textit{I find this system comfortable.} & \cite{borsci22puc} & & \\
         \rog Q14 & \textit{I want to use this system again within the next month.} & \cite{moussawi21em} & & & \\
         Q15 & \textit{I want to use this system regularly over the next few months.} & \cite{moussawi21em} & & & \\

         \bottomrule
    \end{tabular}}

    \label{tab:questions}
\end{table*}

\noindent
\textbf{Post-trial survey.} After the completion of each scenario, participants were asked to complete surveys that assessed their interaction experiences (details provided in \autoref{tab:questions}). For both the CloChat and ChatGPT platforms, we conducted a system-related survey that focused on evaluating the overall quality of the dialogues. This evaluation covered various metrics, including convenience, usefulness, efficacy, overall satisfaction, level of engagement, and the intent to utilize the system in the future. Specifically for CloChat, an additional persona-related survey was conducted. This survey aimed to understand the participants’ experiences with the customized agent personas. It assessed aspects such as perceived empathy, likability, and trustworthiness of the agent personas. The development of the questions for both surveys was informed by an extensive review of academic literature pertaining to persona and conversational agent evaluations (for references, see \autoref{tab:questions}). Participants rated their responses to the survey items on a 7-point Likert scale, with higher scores (closer to 7) indicating a more positive user experience.

\noindent
\textbf{Post-hoc Interview. }
Following the completion of the experimental trials, we engaged participants in semi-structured interviews to delve deeper into their experiences, preferences, and usage of the persona customization feature in CloChat. These interviews were structured around dashboards that summarized key metrics of the study. These metrics included the history of persona customization, conversation logs, and survey results. During the interviews, these dashboard visualizations were collaboratively reviewed with the participants, providing a tangible reference point for discussion. This approach facilitated the generation of insightful follow-up questions, enhancing the depth and relevance of our interviews. On average, each post-hoc interview lasted approximately 18 minutes.

\newcommand{\cg}{\cellcolor{gray!20} }

\begin{table*}[]
\setstretch{1.1}
    \centering
    \caption{List of situational scenarios used in our main study (\autoref{sec:study}). For each situation category (informational, emotional, and appraisal), we generated 1,000 scenarios and picked the representative ones using stratified sampling (\autoref{sec:situations}).  }
    \scalebox{0.95}{
    \begin{tabular}{l|w{l}{9em}w{l}{34em}}
        \toprule
        \textbf{Topic} & \textbf{Situation} & \textbf{Description}\\
        \midrule
         \multirow{7}{*}{\makecell[l]{Informational \\support}}& \makecell[l]{Understanding \\company culture} & \makecell[l]{\textit{Users want to get the feel of a company's culture.} \\ \textit{A conversational agent helps by talking about the basics of corporate culture,} \\ \textit{what's unique to that company, and how to research more about it}}\\
         & \cg \makecell[l]{Handling Stress} & \cg \makecell[l]{\textit{Users look for ways to manage daily stress. A conversational agent shares} \\ \textit{techniques to relieve stress, advice on mental well-being, and other useful resources.}} \\
         & \makecell[l]{Exploring recipes}& \makecell[l]{\textit{Users keen on trying new dishes while chatting with a conversational agent about} \\\textit{cooking methods, ingredients, and handy cooking tips}}\\
         &\cg  \makecell[l]{Travel planning}& \cg \makecell[l]{\textit{Users plan trips by chatting with conversational about preparing for travel,} \\ \textit{cool places to visit, food spots to try, and useful local tips.}}\\
         \midrule
          \multirow{7}{*}{\makecell[l]{Emotional \\support}}& \makecell[l]{Talking About \\ Self-compassion}& \makecell[l]{\textit{When users are too hard on themselves, a conversational agent encourages them} \\ \textit{to be kinder to themselves and offers ways to practice self-esteem}}\\
          & \cg \makecell[l]{Discussions on \\ sleep issues} & \cg \makecell[l]{\textit{Users having trouble sleeping want to talk with a conversational agent} \\ \textit{for understanding and suggestions on how to sleep better}}\\
          & \makecell[l]{Managing nightmares} & \makecell[l]{\textit{For users bothered by bad dreams, a conversational agent offers} \\ \textit{comfort and suggestions on managing them better.}} \\
          & \cg \makecell[l]{Advice on romantic\\  relationships} & \cg \makecell[l]{\textit{Users facing romantic troubles talk with a conversational agent. In response, } \\ \textit{the agent offers understanding and tips for maintaining a healthy relationship.}} \\
          \midrule
          \multirow{7}{*}{\makecell{Appraisal \\support}} & \makecell[l]{Assessing my Skills \\ and personal growth} & \makecell[l]{\textit{Users want to earn feedback on their academic or job skills by chatting with } \\ \textit{conversational agents. Users especially want to figure out strengths,} \\  \textit{areas to work on, and goals for personal growth.}} \\
          & \cg \makecell[l]{Improving problem- \\ solving skills} & \cg \makecell[l]{\textit{Users discuss with a conversational agent how to think} \\  \textit{more logically and make better decisions.}} \\
          & \makecell[l]{Evaluating and building \\ leadership skills} & \makecell[l]{\textit{Users who want to be better leaders discuss leadership styles, effective leadership } \\ \textit{practices, and ways to improve leadership with a conversational agent.}} \\
          & \cg \makecell[l]{Boosting Project \\ management skills} &  \cg \makecell[l]{\textit{Users chat with a conversational agent about how to manage projects better}, \\  \textit{from scheduling to working well with a team.}} \\
          
         \bottomrule
    \end{tabular}
    }

    \label{tab:situations}
\end{table*}

\subsection{Details of Scenarios}
\label{sec:situations}

As detailed in \autoref{sec:procedure}, our study utilized a variety of situational scenarios in which participants engaged with conversational systems. This approach reflects the diverse roles conversational agents play in daily life, as supported by literature \cite{weber2020non}. We adopted Cutrona and Shur's theoretical framework \cite{cutrona1992controllability}, categorizing agents’ social support into three domains: \textit{informational}, \textit{emotional}, and \textit{appraisal support}. Informational support involves providing advice or guidance for everyday challenges \cite{liu2020physician}, emotional support offers empathy and encouragement \cite{flickinger2017social}, and appraisal support aids in self-assessment \cite{cutrona1992controllability}.

To explore these categories, we developed four scenarios for each type of support, totaling 12 distinct scenarios. We employed a stratified sampling method, drawing inspiration from previous studies \cite{pandey2016towards, jeon2023clams}. Initially, we created 10 scenarios for each support category. These scenarios were augmented by ChatGPT (based on GPT-4), which generated 10 additional diverse scenarios. We combined these with our original set and repeated this process 99 times, each time randomly selecting 10 scenarios from the expanded set. This resulted in a corpus of 1,000 scenarios: 10 originally crafted and 990 generated by the model.

The textual descriptions of these scenarios were then transformed into vector embeddings using OpenAI's text-embedding API with the \texttt{text-embedding-ada-002} model. We applied dimensionality reduction to these vectors using the UMATO algorithm \cite{jeon2022uniform}, chosen for its effectiveness in preserving global data structures, in comparison to alternatives like UMAP and $t$-SNE. The effectiveness of this reduction was assessed using Bayesian optimization techniques \cite{snoek2012practical}, with Steadiness \& Cohesiveness as the loss function \cite{jeon22tvcg}.

Finally, we clustered the dimension-reduced vectors using the $K$-Means algorithm, setting $K=4$. We selected scenarios corresponding to the centroids of these clusters for in-depth examination. A complete list of these selected scenarios is available in \autoref{tab:situations}.

\section{Quantitative Results}

We present the quantitative findings of our study. Our initial analysis focused on evaluating the overall user experience and the efficacy of CloChat in comparison to ChatGPT (RQ1). Following this, we explored the methods and patterns with which participants customized their agent personas, as well as their interactions with these personas (RQ2).

\subsection{Analysis of Survey Responses}

\label{sec:quant_survey_sit}

\noindent
\textbf{Objectives.}
Our first analysis aimed to scrutinize and compare the user experiences when interacting with both CloChat and ChatGPT. The focus was particularly on assessing CloChat’s ability to enhance user experience (RQ1). We investigated the differences in the outcomes of post-trial surveys, considering different types of conversational systems and situational contexts.

\noindent
\textbf{Analysis design.}
The survey responses were examined systematically for each question. For system-related attributes, we employed a two-way repeated-measures Analysis of Variance (ANOVA), analyzing the effects of system types (CloChat and ChatGPT) and situational contexts (categorized as informational, emotional, and appraisal support). In the case of the persona-related survey, which pertained to the trials with CloChat, we conducted a one-way repeated-measures ANOVA focusing on the types of situational contexts. To further explore significant findings, we applied Tukey’s Honestly Significant Difference (HSD) test \cite{turkey49biom} for post hoc analysis.

\begin{figure*}
    \centering
    \includegraphics[width=\linewidth]{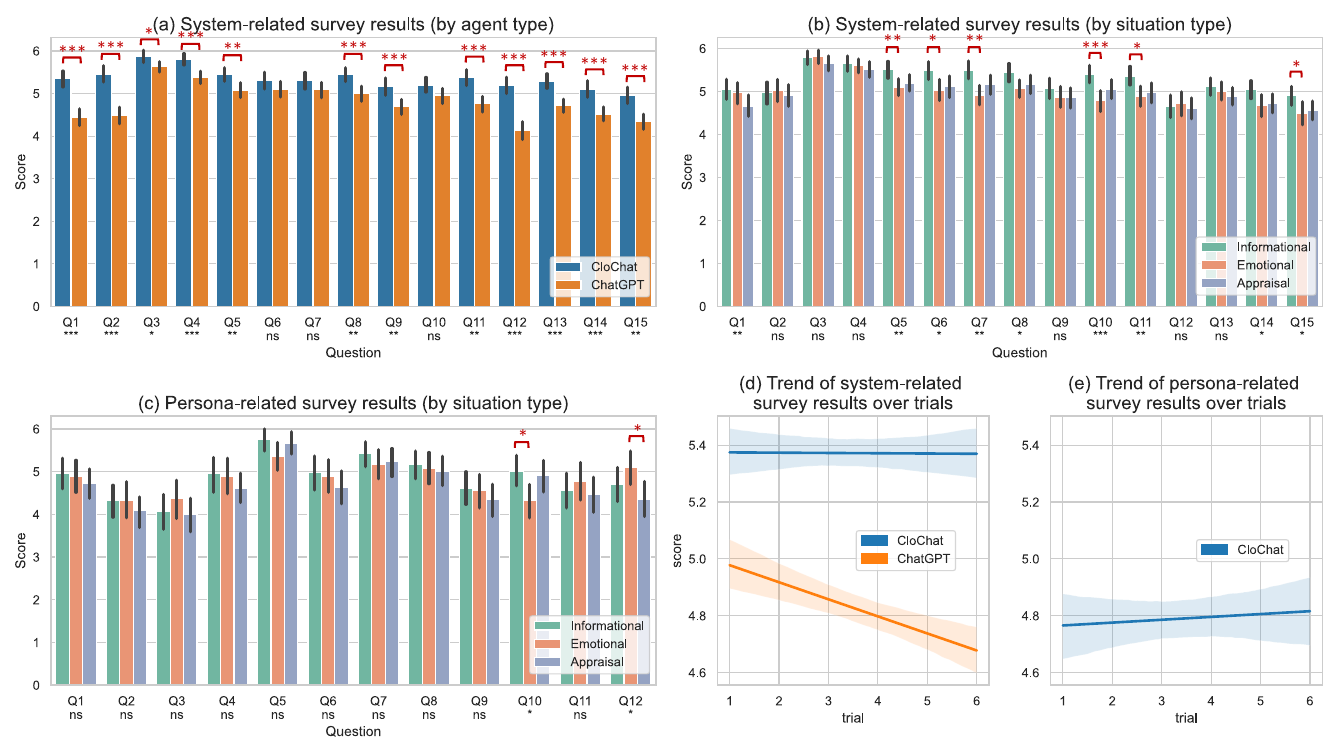}
    \caption{Post-trial survey results (\autoref{sec:quant_survey_sit}, \ref{sec:quant_survey_trials}). a, b) Results of the system-related survey (Table 1a), aggregated by system type and situation type. (c) Results of the persona-related survey (\autoref{tab:questions}b). (d, e) Trends in the system- and persona-related survey scores over trials. For the bar charts (a–c), the asterisks under each question number depict the statistical significance of the repeated-measures analysis of variance values (***: $p < .001$, **: $p < .01$, *: $p < .05$). Statistical significance found in the post hoc analysis is depicted with red brackets. }
    \label{fig:survey}
\end{figure*}

\noindent
\textbf{Results and Discussions.}
The results of our survey are depicted in \autoref{fig:survey} (a-c), and a detailed result of statistical analysis is available in Appendix B. In the system-related survey (questions Q1–Q5, Q8–Q9, and Q11–Q15), we observed a significant main effect related to system types, as shown in \autoref{fig:survey} (a). Our post hoc analysis indicated that CloChat consistently scored higher than ChatGPT across these questions. Although no significant main effects were detected for questions Q6 (\textit{'This system can provide useful answers to me.'}), Q7 (\textit{'This system helps me achieve my goals.'}), and Q10 (\textit{'I feel that this system will make my life more convenient.'}), which focus on the perceived utility of conversational agents (\autoref{tab:questions}), the trend still favored CloChat with higher average ratings.

These findings suggest that CloChat's personalized persona contributes positively to various user experience aspects, such as satisfaction, engagement, and future interaction likelihood. While statistically significant differences in perceived utility items were not observed, a consistent preference for CloChat was evident.

Regarding situation types in the system-related survey, significant main effects were noted for questions Q1, Q5–Q7, Q10–Q11, and Q14–Q15 (\autoref{fig:survey} (b); detailed statistics reported in Appendix B). Post hoc analysis showed that informational situations (Q5–Q7, Q10–Q11, and Q15) garnered higher scores compared to emotional situations, particularly in questions related to system effectiveness, utility, and future use intention. This indicates that users generally perceive conversational agents as more useful and effective for informational support than for emotional support, a trend independent of the presence of personalized personas.

In the persona-related survey, conducted exclusively with CloChat, significant main effects due to situation types were found in Q10 (\textit{'I can utilize this persona for work or academic purposes.'}) and Q12 (\textit{'It feels like this persona has a personality.'}) (\autoref{fig:survey} (c); detailed statistics reported in Appendix B). The post hoc analysis of Q10 revealed significantly higher scores in informational situations than in emotional scenarios, aligning with the question's focus on the persona's utility in academic or professional settings. Conversely, in Q12, emotional situations scored higher than appraisal situations, emphasizing the human-like attributes and emotional resonance of the persona in these contexts.

\subsection{Temporal Evolution of Survey Scores Across Trials}

\label{sec:quant_survey_trials}

\noindent 
\textbf{Objectives.}
The objective of this analysis was to investigate the longitudinal changes in user evaluations of the conversational systems across multiple trials, addressing RQ1-2. Recognizing that higher scores in survey questions could be indicative of a better user experience, we sought to analyze the trends in overall user satisfaction over time.

\noindent
\textbf{Analysis design.}
To visually examine the temporal evolution of survey scores, we conducted regression analyses, plotting distinct regression lines for each conversational agent (CloChat and ChatGPT). For the system-related survey, we utilized an Analysis of Covariance (ANCOVA) \cite{keselman98rer} to statistically assess the significance of the observed differences in score trajectories between the two systems.

\noindent
\textbf{Results and Discussions.}
As depicted in \autoref{fig:survey} (d), survey scores for ChatGPT showed a general downward trend over time, whereas scores for CloChat remained relatively stable. The ANCOVA analysis confirmed that the difference in these trends between CloChat and ChatGPT was statistically significant ($F=89.89$; $p<.001$). Interestingly, the persona-related survey scores, gathered exclusively from CloChat trials, exhibited a slight upward trajectory. These findings suggest that while user experience with conversational agents may typically decline over time, the presence of customized personas in CloChat appears to mitigate this effect, contributing to a sustained or even improved user experience.

\newcommand{\robbb}{\rowcolor{blue!20} }
\newcommand{\robb}{\rowcolor{blue!10} }
\newcommand{\rob}{\rowcolor{blue!5} }

\newcommand{\rorrr}{\rowcolor{red!20} }
\newcommand{\rorr}{\rowcolor{red!10} }
\newcommand{\ror}{\rowcolor{red!5} }

\begin{table*}[]
    \centering
    
    \caption{The categorization of visual traits discovered in our study. Our analysis (\autoref{sec:quant_appear}) shows that traits in \textit{Animals} and \textit{Art \& Styles} categories tend to align less with the persona characteristic compared to the traits in the other categories.}
    \scalebox{0.9}{
    \begin{tabular}{l|l|lc}
    \toprule
       \textbf{Category}  & \textbf{Sub-Category} & \textbf{Example codes} & \textbf{Count}\\
       \midrule
       \multirow{3}{*}{\textit{Animals}} & Cute animals &  Cute cat, Cute puppy, Cute bear, Cute panda, Cute seal & 17\\
                                &  Specific breeds & Korean Shorthair cat, Golden Retriever, Border Collie &  3 \\
                                & Animal behavior or moods & Wagging tail, Smile & 3 \\
        \rog  & Asian influences& Korean, East Asian woman, Vietnamese merchant& 10 \\
                                                          \rog  \multirow{-2}{*}{\textit{Cultural or Regional Traits}}  & Western influences
             & British gentleman, White Western woman &  3\\
        \multirow{4}{*}{\textit{Professions \& Roles}} & Business-related roles & Company employee, Executive in a startup, Office worker & 6\\
                                                      & Academic professions & Professor, Graduate Student&  5\\
                                                      & Service roles & Chef, Butler, Counselor, Doctor, Guide &  7\\     
                                                      & Creative roles  & YouTuber, Actress, Musician &  3\\
        \rog  & Hair types \& styles & Short hair, Perm, Bald, Beard & 9\\
                                                         \rog         & Clothing \& accessories & Suit, Doctor's gown, Hat, Glasses, Pajamas, Hawaiian shirt & 9 \\
                                                          \rog        & Age & Baby, Middle-aged, in their 20s/30s/40s/50s &  9\\     
                                                         \rog \multirow{-4}{*}{\textit{Detailed Physical Appearances}}  & Expressions and demeanor  & Smiling, Serious, Tough, Friendly,&  4\\           

        \multirow{3}{*}{\textit{Art \& Style}} & Painting styles & 2D, 3D, Oil painting, Disney style & 8\\
                                                                  & Settings \& background & Amusement park, Forest, Office&  3\\
                                                                  & Mood or emotion & Bright, Cute, Comfortable, Mysterious&  4\\     
        \rog  & Quirky ideas & Cyber Buddha, Virgin Mary with electric guitar, Zhuangzi with wine & 3\\
                                                   \rog               & Descriptive  traits& Diligent, Hardworking, Charismatic, Kind &  9\\
                                                     \rog  \multirow{-3}{*}{\textit{Unique and Abstract Concepts}}           & Non-human characters & Rock, Blue square, Ghost &  3\\   
        \bottomrule
    \end{tabular}}
    \label{tab:visual_representation}
\end{table*}

\begin{figure}
    \centering
    \includegraphics[width=\linewidth]{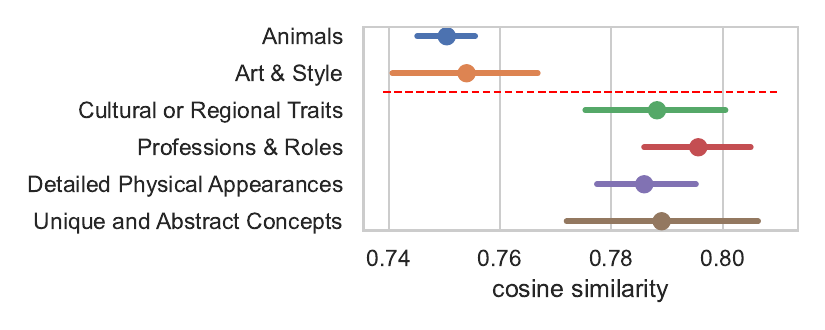}
    \vspace{-4mm}
    \caption{Degree of alignment between persona characteristics (non-visual traits) and visual traits based on the category of visual traits (\autoref{tab:visual_representation}). Higher cosine similarity scores represent better alignment. Post hoc analysis revealed that the categories on the upper side of the red dashed line obtained significantly lower cosine similarity scores than those on the lower side. }
    \label{fig:visual_category_sim}
\end{figure}

\subsection{Alignment between the Visual and Non-visual Traits of Agent Personas}

\label{sec:quant_appear}

\noindent
\textbf{Objectives.}
In addressing RQ2 in a detailed way, we aimed to examine the correlation between the visual representations and non-visual characteristics (such as personality traits and roles) of customized agent personas. Existing literature suggests that a conversational agent's visual appearance often correlates with its persona, where specific traits or roles influence its visual depiction \cite{nguyen22idc, purington17chiea, wang12iconf}. Our goal was to delve into this alignment, exploring how individuals intentionally coordinate these visual elements with their personalized agent personas.

\noindent
\textbf{Analysis design.}
We began with axial coding to categorize the relationship between various traits and the visual representations of agent personas. Two researchers independently created codebooks, which were then merged after discussions for consistent analysis.

To understand how the relationship between visual and non-visual traits varies across different visual trait categories, we first identified agent personas from our study where the visual representation fell into specific categories. 

Next, we converted the visual and non-visual traits of these agent personas into vector embeddings. For visual traits, we used OpenAI’s text-embedding API (with the \texttt{text-embedding-ada-002} model) to transform image prompts into vectors. For personality traits, we transformed the natural language directives used in GPT-4 (refer to \autoref{sec:architecture}) into vector embeddings. We then calculated the cosine similarity between the vectors representing the image prompts and those representing the persona characteristics. A one-way ANOVA was conducted to assess differences in similarity scores across categories, followed by a post hoc analysis using Tukey’s HSD test.

\noindent
\textbf{Results and discussions.}
Our analysis yielded six distinct categories of traits associated with visual representation:  \textit{Animals}, \textit{Cultural or Regional Traits}, \textit{Professions \& Roles}, \textit{Detailed Physical Appearances}, \textit{Art \& Style}, and \textit{Unique \& Abstract Concepts} (for detailed coding results, see \autoref{tab:visual_representation}). A one-way ANOVA revealed significant differences in cosine similarity scores among these categories ($F(5, 148) = 8.190$, $p < .001$). Post hoc analysis using Tukey’s HSD identified notably lower scores in the \textit{Animals} and \textit{Art \& Style} categories compared to the others (\autoref{fig:visual_category_sim}). For more detailed statistical information ($p$-values and confidence intervals), please see Appendix B.

A key distinction between categories with high and low similarity scores is the direct relevance of visual traits to human characteristics. 
Categories such as \textit{Professions \& Roles} (including specific roles like Office Worker, Professor, YouTuber)and \textit{Cultural or Regional Traits} category (e.g., Korean, British) explicitly denote human subgroups, while the \textit{Detailed Physical Appearances} category focuses on human features. Similarly, the \textit{Unique and Abstract Concepts} category generally relates to human attributes, barring some non-human focused subcategories. In contrast, the \textit{Art \& Style} and \textit{Animals} categories predominantly include traits that do not directly correspond to human attributes.

The results indicate that when participants chose visual traits closely linked to real-world human characteristics for their agent personas, there was a greater likelihood of alignment between these visual elements and the agent personas’ non-visual traits. This tendency might also suggest that users often perceive their agent personas as virtual humans, expecting them to visually mirror typical human characteristics. Conversely, traits not directly related to human attributes tend to be applied more flexibly, reflecting individual user preferences rather than a strict alignment with the non-visual traits of their agent personas.

\subsection{Diversity of Dialogues}

\label{sec:diversity}

\begin{figure}
    \centering
    \includegraphics[width=\linewidth]{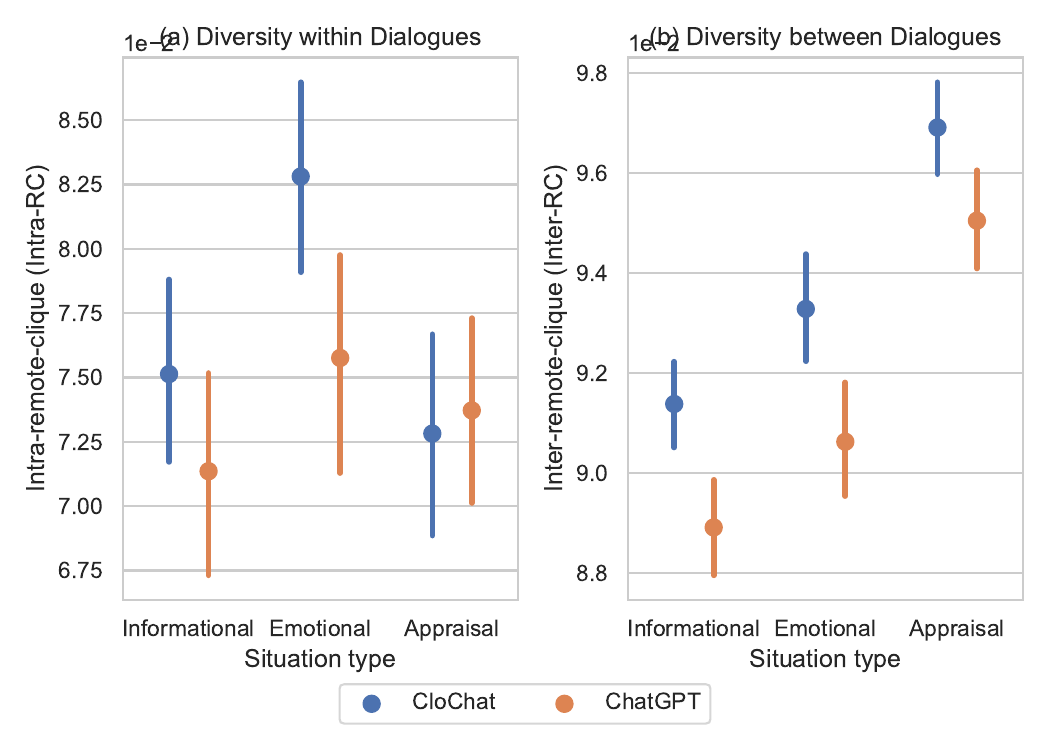}
    \caption{Diversity of dialogues created in our study, assessed by inter- and intra-remote-clique measures (\autoref{sec:diversity}). 
    In summary, the dialogues with CloChat showed substantially higher diversity than those with ChatGPT. }
    \label{fig:diversity}
\end{figure}

\noindent
\textbf{Objectives.}
In relation to RQ2, we hypothesized that using CloChat would lead to more enriched and diverse dialogues with conversational agents compared to standard ChatGPT interactions. Our goal was to empirically validate this hypothesis and explore the impact of different situational scenarios on the diversity of dialogues.

\noindent
\textbf{Analysis design.}
To rigorously evaluate dialogue diversity, we developed two specialized metrics: intra-remote-clique (intra-RC) and inter-remote-clique (inter-RC). These metrics are adaptations of the remote-clique (RC) metric \cite{cox21chi}, which is commonly used to measure text embedding diversity. The RC metric is defined as the average pairwise distance between text embeddings \cite{dow11tochi, kaminskas15tiis}. Intra-RC specifically measures the average pairwise distance between utterances within a single dialogue, providing insight into the diversity of conversation within one session. Inter-RC, on the other hand, assesses the average linkage between utterances across two dialogues within the same situational context, offering a perspective on the diversity between different conversations under similar circumstances.

For each dialogue in our study, we computed the intra-RC to determine the level of diversity within that dialogue. We also calculated the inter-RC for each pair of dialogues sharing the same situational context to evaluate the diversity between conversations. To ensure that our metrics were not influenced by the semantic differences between various scenarios, we avoided comparing dialogues from distinct scenarios. We then conducted a two-way ANOVA to analyze the effects of system type (CloChat and ChatGPT) and situation type (informational, emotional, and appraisal support) on dialogue diversity. Tukey’s HSD test was carried out for the post hoc analysis.

\noindent
\textbf{Results and discussions.}
The findings from our analysis are depicted in \autoref{fig:diversity}. 
In terms of post hoc analysis, please refer to Appendix B.
For intra-RC, a significant main effect was observed for system types ($F(1,354) = 4.16$, $p < .05$). However, post hoc analyses did not reveal any statistically significant differences between CloChat and ChatGPT. Regarding situation types, a significant main effect was also noted ($F(2, 354) = 6.13$, $p < .01$). Post hoc tests showed that dialogues in emotional contexts exhibited significantly higher diversity compared to both informational ($p < .01$) and appraisal ($p < .01$) contexts. We did not identify any interaction effects between system and situation types.

In the case of inter-RC, there were notable main effects for both system types ($F(1,5214) = 30.91$, $p < .001$) and situation types ($F(2,5214) = 67.71$, $p < .001$). Post hoc analysis revealed that dialogues using CloChat displayed a significantly higher level of diversity compared to ChatGPT ($p < .001$). Furthermore, we observed a systematic increase in dialogue diversity across the informational, emotional, and appraisal scenarios, with statistically significant differences in all pairwise comparisons ($p < .001$ for each). Again, no interaction effects were found.

To summarize, the results indicate that CloChat significantly enhanced the diversity of dialogues between different conversations (inter-dialogue diversity), but did not have a marked effect on the diversity within individual conversations (intra-dialogue diversity), in comparison to standard ChatGPT interactions. This suggests that while CloChat's tailored agent personas contribute to personalizing conversations, they may not necessarily increase the dynamic range of topics or conversational patterns within a single dialogue session.

\section{Qualitative Results}

In addition to our quantitative analysis, we delved into qualitative data from interviews to gain deeper insights into our research questions. We employed thematic analysis \cite{braun2012thematic} as our methodological framework for the analysis. The research team utilized a line-by-line open coding technique, allowing for the identification and categorization of emergent themes from the interview data. The findings from this thematic analysis are detailed in the subsequent sections.

\subsection{Patterns in Customizing and Selecting Agent Personas}
\label{sec:qual_pattens}

Our user study revealed two distinct patterns in the creation and reuse of agent personas, each illustrating unique approaches to user engagement and satisfaction. The first pattern is characterized by dynamic persona customization, specifically tailored to meet immediate situational needs. Participants following this approach proactively envisioned specific scenarios for interaction and selected personas with appropriate characteristics, like personality and expertise, to match these situations. On average, participants in this group changed their agent personas 4.6 times over the six trials with CloChat, with more than half using six different personas for each session. For example, Participant 20 created a ‘psychiatrist’ persona to address stress and sleep concerns, commenting, \textit{“I was super stressed, so I thought, why not talk to a 'psychiatrist'?”} Similarly, in career guidance scenarios, participants customized personas to mimic employees from companies of interest, reflecting the importance of contextually relevant and personalized conversational experiences.

Conversely, the second pattern indicates a preference for reusing specific agent personas that have previously provided satisfactory conversational experiences. In our study, 12 participants consistently reused a particular persona for more than two trials, with some using the same persona throughout all six trials. For example, P11 repeatedly chose the ‘gentleman persona,’ stating, \textit{“I kept using the 'gentleman' because he just gets me. He always knows the right thing to say.”} This pattern suggests that once a persona resonates with a user's expectations, it fosters a sense of trust, reinforcing the user's initial choice and encouraging future interactions. P27, for example, continued using a persona initially selected on a whim due to its unexpectedly accurate responses, saying, \textit{“At first, I picked the persona just for kicks. But it was so on point, I kept coming back.”}

These two patterns differ fundamentally in their approach: the first is dynamic, with participants varying persona characteristics to suit specific needs, while the second is consistent, favoring a particular persona based on personal satisfaction and preference. This dichotomy illustrates how individual user preferences and needs can manifest in diverse ways when engaging with conversational systems, balancing between situational diversity and consistent personal preferences.

\subsection{Conversation Diversity and Dynamics}
\label{sec:qual_Diversity}

The study revealed that the use of agent personas in conversational agents can offer a more diverse and enriched dialogue experience for participants. Initially, some participants expressed during interviews that they primarily utilized LLMs for basic tasks like answering simple questions or conducting fundamental information searches, valuing ChatGPT's immediate response capabilities over complex customization options. However, post-experiment interviews revealed a notable shift in perception. P8 observed, \textit{“Even if the answers are the same, having a persona adds a more professional feel. I think it could be useful even in casual conversations.”} The comment suggests that customized agent personas can influence their user experience in a positive way, indicating a potential shift in user behavior from basic information retrieval to seeking more personalized and engaging interactions.

Despite the experiment's scenarios being categorized as informational, emotional, and appraisal, participants often ventured beyond these confines. Their intrigue with personalized agent personas led them to explore new topics and questions. P13 reflected, \textit{“The conversation got longer when I found more fun and interesting topics, similar to talking with friends. With ChatGPT, the conversations were shorter due to predictable responses.”} This expansion in dialogue scope fostered deeper and more intricate relationships between participants and agent personas. 

The agent personas not only influenced the nature of the dialogue but also affected the participants’ conversational styles. Engaging with a friendly and humorous persona, for example, fostered a light-hearted atmosphere, encouraging participants to use informal language and share jokes. P27 noted, \textit{“Talking to this persona felt like chatting with an old friend. I often found myself laughing.”} Conversely, interactions with more serious or formal personas led to dialogues with a scholarly or cautious tone. P13 commented, \textit{“My persona was cold and academic, like Sherlock Holmes, which naturally steered the conversation to be more serious.”} These dynamics even impacted the participants' moods and emotions, as highlighted by P30: \textit{“I felt more energetic talking to my vibrant persona, whereas serious conversations prompted deeper thought.”}

\subsection{Relationship between Participants and Agent Personas}

\label{sec:qual_relation}
The introduction of agent personas led participants to perceive their conversational partners as entities with unique contexts and personalities, rather than just as programs. Many participants reported enhanced immersion and trust in their interactions when the agent persona's responses aligned with their expectations or preferences. For example, P1 expressed, \textit{“Having a personalized persona made the conversation feel more alive, and I felt more trust in the interaction.”} This increase in trust, as evidenced by a previous study \cite{lessio2020toward}, highlights the importance of persona alignment in fostering meaningful conversational experiences.
 
In contrast, interactions with ChatGPT were often perceived as engaging with an automated responder, lacking a personal touch. Participants like P13 remarked, \textit{“My conversations with ChatGPT felt pretty standard. It was like getting necessary information from a machine, without any specific expectation or connection.”} This difference underscores the uniqueness and personalization that agent personas can bring to conversational experiences.

Another significant aspect of our findings pertains to the emotional connection participants developed with their configured personas. Some participants experienced profound emotional responses during their interactions. P6 shared, \textit{“The conversation moved me almost to tears,”} while P19 described the conversation as akin to talking with a friend due to the persona's empathy. 

The visual representation of personas also played a crucial role in enhancing empathy and engagement. P15 mentioned, \textit{“Seeing the persona I created made the conversation feel more direct, eliciting a deeper sense of empathy.”} The act of visualizing and personalizing these personas enriched the conversational experience, as P9’s comment illustrates: \textit{“I crafted it thinking of my favorite YouTuber. During our chat, I imagined his voice, making the conversation more engaging.”} This aligns with research findings that emphasize the power of visual engagement in enhancing conversational interest \cite{University2023}.

Initially, many participants were not inclined to use conversational agents for emotional support, a trend also supported by our quantitative findings (\autoref{sec:quant_survey_sit}). However, as the experiment progressed, participants began to appreciate the value of emotional conversations with agent personas. P19’s reflection captures this shift: \textit{“The experiment taught me the value of emotional conversations with conversational agents. CloChat’s personalized agents responded warmly, understanding my feelings remarkably well.”}

\subsection{User Feedback on the Persona Customization}

\label{sec:qual_burden}

The participants found that CloChat’s form-based interface significantly lowered the entry barrier for engaging with conversational agents, making it more accessible to the general public. During pre-experiment interviews, many participants revealed difficulties due to limited technical knowledge needed for LLM customization, particularly when it came to selecting specific characteristics for personas. 
Thus, for participants unfamiliar with crafting text prompts, the availability of predefined persona trait options in CloChat was notably more user-friendly. P17, who had initially been concerned about the complexity of prompt creation, observed after the experiment, \textit{“CloChat definitely reduces the effort needed to create a persona. It’s convenient not having to think about specific text prompts.”} As the trials progressed, participants developed their own strategies for effectively customizing unique personas. P30 commented, \textit{“Customizing personas was initially challenging, but I quickly discovered the optimal approach.”} This feedback indicates that users experienced a manageable learning curve with the CloChat interface.

Nevertheless, some participants pointed out that setting up personas could be complicated and time-consuming without clear guidelines or presets. P15 noted, \textit{“I was a bit confused when first setting up the persona. I wasn't sure how to approach it or what criteria to use for selection.”} While most acknowledged the benefits of having bespoke personas, there were mentions of the burden involved in their initial setup as well, suggesting a need for more user-friendly guidance or preset options.

The feature allowing users to customize visual representations of agent personas was particularly appreciated, offering an enhancement not found in ChatGPT. P13 remarked, \textit{“Modifying the counselor’s appearance was surprising and greatly enhanced my engagement.”} This emphasizes the vital role of visual representation in the design and functionality of conversational agents, enhancing user engagement and expectation management.

\subsection{Reflecting Real Life to Agent Personas}

\label{sec:qual_reflect}

A notable trend among participants was the incorporation of elements from their real-life experiences and observations into their agent personas, rather than creating entirely fictional characters. For instance, participants often modeled personas after familiar individuals like acquaintances, friends, pets, or celebrities. P9, who chose a renowned doctor as a persona, shared, \textit{“I based the persona on a real person I saw on TV. Reflecting his tone in my agent persona made the conversation warmer and more immersive, allowing me to speak more honestly.”} Similarly, P27 created a persona inspired by a friend’s occupation and hobbies, noting, \textit{“Seeing these characteristics in the conversation gave it the feeling of talking to my actual friend.”} This approach illustrates how personal experiences can enhance the realism and relatability of conversational partners. However, this practice can also raise ethical concerns regarding privacy and personal data protection, as it involves imitating or mimicking real individuals potentially without their consent.

Participants also enjoyed the imaginative exercise of setting up their pets as personas, attributing them with imagined personality traits and habits. P2 reflected, \textit{“I mirrored my dog's playful personality. Imagining his responses made the conversation more fun and unique.”}

The practice of drawing from real-life experiences for persona customization allowed participants to infuse their personal lives and emotional connections into the digital domain. Nevertheless, while this approach significantly enriches user interaction with conversational AI systems, it simultaneously highlights the importance of addressing ethical considerations related to mimicking real-world individuals.

\section{Discussions}
Our user study was aimed at investigating the impact of agent persona customization on user experience during interactions with LLM-based conversational agents, as opposed to conventional generic conversational agents (RQ1). We discovered that the customization of agent personas significantly boosts user engagement, trust, and emotional connection, offering a noticeable improvement in maintaining user satisfaction and engagement compared to ChatGPT. In addressing RQ2, we delved into the ways users customize their agent personas and the resultant effects on their interactions. We observed that conversations involving customized agent personas tend to be richer and more diverse. Users often align the traits of agent personas in terms of both visual elements and real-world inspirations, which additionally brings to light ethical considerations regarding agent persona customization. In extending our discussions on these findings, we explore relevant topics and present practical implications for the design of user interfaces employing LLM-based conversational agents. We also outline the limitations of our study, acknowledging areas that could benefit from further exploration and improvement.

\subsection{The Multifaceted Roles of Customizable Agent Personas}

\label{sec:disc_enhancing}

Our study demonstrated that CloChat provided an enhanced user experience compared to ChatGPT, highlighting the substantial benefits and potential of customizable agent personas. Users interacting with CloChat perceived the agent personas not just as algorithmic tools, but as distinct conversational partners with unique personalities, as outlined in (\autoref{sec:qual_relation}). This shift in perception, supported by previous research \cite{lessio2020toward,moussawi2021effect,lee2019does}, increased users' emotional engagement, trust, and immersion in the conversational experience.

A noteworthy observation was how some participants modified their own conversational styles to resonate more with the personas they created, indicating a deepening emotional connection with their customized agents (\autoref{sec:qual_relation}). The integration of visual representations further solidified this bond, elevating the agents from mere information retrieval tools to authentic conversational partners (\autoref{sec:qual_relation}, \autoref{sec:quant_appear}).

Conversely, interactions with ChatGPT were associated with lower levels of emotional engagement (\autoref{sec:qual_relation}). This contrast not only underscores the limitations of text-prompt-focused platforms like ChatGPT but also highlights the potential of CloChat's comprehensive personalization features. These features can have the ability to enrich user experiences across diverse emotional contexts and situations.

In conclusion, the customizable agent personas in CloChat extend beyond traditional information retrieval roles typically associated with conversational agents using LLMs. They play a crucial role in fostering emotional connections and enhancing user engagement with conversational systems, indicating an expansion in both the functional scope and emotional depth of these technologies.

\subsection{Personas' Role in Sustaining User Engagement on Conversational Agents}
\label{sec:disc_sustaining}

While ChatGPT is renowned for its conversational capabilities, it faces limitations in reflecting users’ individual preferences and sustaining deep, ongoing relationships, as it primarily excels in basic information retrieval and short interactions \cite{brandtzaeg2018chatbots}. Our study confirms this, indicating a decline in user satisfaction with ChatGPT over time (\autoref{sec:quant_survey_trials}).

In contrast, personalized agent personas not only elicited initial positive responses from users but also played a pivotal role in maintaining these positive connections over time (\autoref{sec:quant_survey_trials}). This aligns with prior research \cite{braun2019your,volkel2021eliciting} and our qualitative findings (\autoref{sec:qual_pattens}), suggesting that user preferences are dynamic, varying according to mood, situation, and context \cite{book}. CloChat’s capability to customize a variety of personas to adapt to these shifting preferences likely contributed to sustained user engagement.

Another key factor in the enduring positive relationship with personalized agent personas is the human-like perception they create (\autoref{sec:qual_relation}), resonating with findings from Cowan et al. \cite{cowan2017can}. With CloChat, participants engaged in longer conversations and explored a wider range of topics (\autoref{sec:qual_Diversity}, \ref{sec:diversity}), leading to increased trust and satisfaction. This enriched conversational experience contributes to sustainable interaction with the agent, moving beyond brief, transactional conversations. Although our study did not specifically observe long-term interactions between users and customized agents, the implications from our findings hint at the potential for fostering lasting relationships with conversational agents in the future.

\subsection{Pros and Cons of Persona Customization}

\label{sec:disc_procon}

Our study underscores the significant advantages of incorporating persona customization features into LLM-based conversational user interfaces. The majority of participants responded positively to this functionality, noting that it made their conversations more enjoyable and engaging (\autoref{sec:qual_relation}, \autoref{sec:qual_burden}). The ability to tailor personas according to personal preferences fostered increased interest and active participation in conversations, leading to a more open and dynamic interaction, as reflected in survey responses (\autoref{sec:quant_survey_sit}).

However, alongside these benefits, certain challenges were also observed (\autoref{sec:qual_burden}). Some participants found the wide array of customization options to be overwhelming, particularly for those new to conversational agents or not versed in prompt engineering techniques. To address this, future iterations could consider integrating automated suggestions that assist users in managing their expectations and simplifying the decision-making process. This could involve methods like OpenAI’s recently released GPTs, which can learn specific knowledge or personalities from user-provided documents~\cite{gpts}. Further research is needed to compare various approaches, such as extensive user-driven customization versus agents automatically learning from user documents, and to understand how these different methods influence user experience. An effective balance between user-driven customization and automated recommendations, as suggested in literature \cite{knijnenburg2012explaining}, could provide a solution to these challenges.

The overarching aim would be to streamline the customization process, making it less daunting for users while still offering a rich, personalized experience. This balance is key to harnessing the full potential of persona customization in enhancing user engagement with conversational AI systems.

\subsection{Ethical Concerns on Personalized Personas}

\label{sec:disc_ethical}

In our study, we observed that participants frequently drew inspiration from their personal experiences and daily interactions when customizing their agent personas (see \autoref{sec:qual_reflect}). A notable trend involved mimicking celebrities or personal acquaintances. This inclination could be attributed to the perceived expertise or symbolic stature of famous individuals or a preference for replicating interactions with familiar and relatable figures rather than inventing entirely new or unknown personas. While this method can lend a sense of realism to interactions with conversational agents and potentially foster more robust and lasting connections, it also brings forth significant ethical dilemmas.

This practice might risk privacy breaches and confidentiality issues, particularly when integrating distinct details or characteristics of these individuals, such as their occupation, location and relationships with others. Furthermore, since a persona cannot fully encompass the complexity of an actual person's personality, actions, or thoughts, such representations may lead to misconceptions or biases. These misrepresentations could adversely impact the reputations or identities of the individuals portrayed, as discussed in the research by Deshpande et al. \cite{ameet2023toxicity}. Hence, a delicate balance must be struck between the creative liberty in persona customization and the ethical implications of drawing from real-life figures.

In the context of LLMs operating across networks, using personal information to shape agent personas raises concerns about individual privacy. Once personal identifying data is input into an LLM, its permanence and the opaque nature of data storage and processing can result in unintended privacy violations, with interactions potentially reaching a broad, unknown audience.

Echoing the observations of Goldstein et al. \cite{goldstein2023generative}, the realm of AI ethics is continuously evolving. Ongoing dialogue and development are essential to establish ethical frameworks and principles within this field. Consequently, it's critical to develop practical and robust solutions for ethical issues related to language model applications. Researchers and developers should diligently address these ethical aspects in the design and deployment of personas, implementing safeguards to protect personal information during the training of machine learning models. This step is fundamental to preserving user privacy and ensuring the ethical use of LLM-based conversational systems. Moreover, clear ethical guidelines and protocols for persona design are necessary. Users should be informed about the risks of imitating real individuals and discouraged from engaging in such practices. Future research should delve into the potential problems of using personas based on real individuals in specific contexts. It may be advisable to limit the use of personas based on real people, especially in scenarios requiring expert advice or sensitive discussions (e.g., sexual dialogue). Such measures will help users grasp the ethical implications of their choices and encourage responsible persona creation.

\subsection{Design Implications}

\label{sec:implications}

Based on our discussions, we propose the following design implications for future development and refinement of conversational user interfaces employing LLMs:

\begin{itemize}
    \item \textbf{I1: Prioritize Persona Customization to Enhance User Trust and Engagement.} CloChat surpassed ChatGPT in terms of user satisfaction, largely owing to the availability of customizable personas. Designers should, therefore, consider prioritizing persona customization options in their systems. The heightened user trust and improved conversation quality associated with personalized personas highlight their vital role in the future design of conversational interfaces.
    
    \item \textbf{I2: Minimize the Initial Setup Burden to Encourage User Engagement.} The initial setup for persona customization can be perceived as burdensome (\autoref{sec:disc_procon}). Designers should streamline the setup process and provide easy-to-follow onboarding assistance, thus enhancing user immersion and engagement from the outset (\autoref{sec:disc_enhancing}, \ref{sec:disc_sustaining}). This could involve introducing conversational tutorials or preset persona options.
    
    \item \textbf{I3: Make the Agent Persona Adaptive.} The flexibility in creating bespoke agent personas for various situations could be helpful for sustaining user engagement with CloChat. We recommend implementing adaptive algorithms that tailor persona behaviors based on user intentions and circumstances, combined with easy customization options. This approach would cater to users who prefer consistent personas across different scenarios as well as those who desire situation-specific persona adaptations.
    
    \item \textbf{I4: Provide Thorough Guidelines on Ethical Considerations.} The study revealed potential ethical issues, such as incorporating celebrities or real-life acquaintances into agent persona designs without consent. Given the likelihood of further ethical concerns, it is crucial to provide users with clear guidelines addressing these issues. This will help ensure that the customization of agent personas adheres to ethical standards and respects individual privacy and rights.
\end{itemize}

\subsection{Limitations and Future Work}
Our study, while shedding light on the diverse user experiences with customizable agent personas in LLMs, has several limitations that must be acknowledged. Firstly, the participant pool was limited to Korean speakers, due to institutional constraints. This limitation may affect the generalizability of our findings to other linguistic and cultural groups. Future studies should aim to include a more diverse range of participants to broaden the applicability of the results.  
Secondly, the creation of agent personas in CloChat relied solely on prompt injection, which may lack depth in specialized or rapidly evolving domains \cite{agarwal2023kitlm}. This limitation raises the question of how LLMs can be optimized for more in-depth and accurate persona representations. Future research could explore advanced techniques such as fine-tuning \cite{nikolich21dsis} or the integration of external memory \cite{park2023generative, schuurmans2023memory} to enhance the sophistication of persona customization in LLMs. Thirdly, CloChat's persona customization is currently confined to a form-based interface. While this design choice was made to lower the barrier to persona customization, it is worth exploring how different customization methods (e.g., direct prompt writing, conversation-based customization \cite{introducinggpts}) might impact user experience. These alternative approaches could offer more flexibility and personalization, catering to users with varying levels of expertise and preferences.
Lastly, our study did not explore the long-term user experience with CloChat. To fill this gap, future research endeavors should focus on longitudinal studies to understand how user engagement with CloChat evolves over time. Such studies are crucial for uncovering the distinctions between short-term and long-term interactions and for developing strategies to cultivate sustained and meaningful relationships with conversational agents like CloChat.

\section{Conclusion}

In this study, we explored how users engage with and customize agent personas in LLM-based user interfaces. For this purpose, we developed CloChat, a user-centric interface built upon ChatGPT, enabling users to easily customize and interact with agent personas. We then compared CloChat with the standard ChatGPT system through a user study. Our findings indicate that CloChat significantly improves the overall user experience compared to ChatGPT, suggesting that giving users the ability to personalize their conversational agents leads to a more satisfying experience. Additionally, it was observed that users not only enjoy the process of customizing their agents but also find these personalized agents to be more engaging conversational partners. Users developed more meaningful relationships with the customized personas and engaged in more prolonged interactions with them. Drawing from these insights, we proposed design implications for future systems utilizing conversational agents. We hope this will pave the way for groundbreaking advancements in the design of conversational interactions facilitated by LLMs.

\section*{Acknowledgments}

This research was supported by the Yonsei University Research Fund of 2023 (2023-22-0430) and by the National Research Foundation of Korea (NRF) grant funded by the Korea government (MSIT) (No. 2023R1A2C200520911). 
This  work  was  also supported  by the Institute  of  Information  \&  communications 
Technology  Planning  \&  Evaluation  (IITP)  grant  funded  by  the  Korea 
government (MSIT)  [NO.2021-0-01343,  Artificial  Intelligence  Graduate  School 
Program  (Seoul  National  University)]
The ICT at Seoul National University provided research facilities for this study.

\bibliographystyle{ACM-Reference-Format}
\bibliography{mainref}










\end{document}